\documentclass[review]{elsarticle}

\usepackage{amssymb}
\usepackage{amsmath}
\usepackage{subfigure}
\usepackage{bm}
\usepackage{booktabs}
\usepackage[usenames,dvipsnames]{color}

\journal{Chaos, Solitons and Fractals}

\begin{document}

\begin{frontmatter}



\title{Stable chaos in fluctuation driven neural circuits}


\author[label1]{David Angulo-Garcia\corref{cor1}}
\ead{david.angulo@fi.isc.cnr.it}
\author[label1,label2]{Alessandro Torcini}
\ead{alessandro.torcini@cnr.it}

\address[label1]{CNR - Consiglio Nazionale delle Ricerche -
Istituto dei Sistemi Complessi, via Madonna del Piano 10,
I-50019 Sesto Fiorentino, Italy}

\address[label2]{INFN Sez. Firenze, via Sansone, 1 - I-50019 Sesto Fiorentino, Italy}

\cortext[cor1]{Corresponding author at: CNR - Consiglio Nazionale delle Ricerche -
Istituto dei Sistemi Complessi, via Madonna del Piano 10,
I-50019 Sesto Fiorentino, Italy. Tel.: +39 055 522 6678.}

\begin{abstract}
We study the dynamical stability of pulse coupled networks of leaky integrate-and-fire 
neurons against infinitesimal and finite perturbations. In particular, we compare current 
versus fluctuations driven networks, the former (latter) is realized by considering purely excitatory
(inhibitory) sparse neural circuits. In the excitatory case the instabilities of the system
can be completely captured by an usual linear stability (Lyapunov) analysis,
on the other hand the inhibitory networks can display the coexistence of
linear and nonlinear instabilities. The nonlinear effects are associated to
finite amplitude instabilities, which have been characterized in terms of 
suitable indicators. For inhibitory coupling one observes a transition
from chaotic to non chaotic dynamics by decreasing the pulse width.
For sufficiently fast synapses the system, despite showing an erratic 
evolution, is linearly stable, thus representing a prototypical example of 
{\it Stable Chaos}.
\end{abstract}


\end{frontmatter}



\section{Introduction}

It is known that cortical neurons {\it in vivo} present a high
discharge variability, even if stimulated by current injection, in comparison
with neurons {\it in vitro} ~\cite{SoftkyKoch1993,holt1996}. In particular,
these differences are peculiar of pyramidal neurons, while inter-neurons
reveal a high neuronal firing variability in both settings~\cite{Stiefel_Sejnowski2013OriginIrregular}.
This variability is usually measured
in terms of the coefficient of variation $CV$ of the single neuron inter-spike interval (ISI),
defined as the normalized standard deviation of the ISI, i.e, $CV = STD(ISI)/\langle ISI\rangle$
\cite{tuckwell2005}. For cortical pyramidal neurons $CV \simeq 1.0$ {\it in vivo}~\cite{SoftkyKoch1993}
and $CV < 0.3$ {\it in vitro}~\cite{holt1996}, while for cortical inter-neurons  $CV \simeq 1.0 - 1.2$
~\cite{Stiefel_Sejnowski2013OriginIrregular} in both settings. The variability of the spike
emissions {\it in vivo} resembles a stochastic (Poissonian) process (where $CV=1$), however 
the neural dynamics features cannot 
be accounted by simple stochastic models \cite{SoftkyKoch1993}. 
These phenomena can be instead modelized by considering a deterministically
balanced network, where inhibitory and excitatory activity on average 
compensate one another~\cite{Abbott1993Asynchronous,Haider2006BalancedExperiment,renart2010,litwin2012slow}. Despite 
the many papers devoted in the last two decades to this subject, is still unclear
which is the dynamical phenomenon responsible for the observed irregular 
dynamics~\cite{TsodyksSenojwski1995,BrunelHakim1999,NeuronalNetwroks2005Vogels,Brunel2000Sparse}.

A few authors pointed out the possibility that {\it Stable Chaos}~\cite{politi2010stable}
could be intimately related to the dynamical behavior of balanced 
states~\cite{Zillmer2006,JhankeTimme2008PRL,Zillmer2009LongTrans,Timme2009ChaosBalance,Luccioli2010Irregular,MonteforteBalanced2012}. Stable Chaos is a dynamical regime characterized by linear stability
(i.e. the maximal Lyapunov exponent is negative), yet displaying an erratic behavior
over time scales diverging exponentially with the system size.
Stable Chaos has been discovered in arrays of diffusively coupled discontinuous maps~\cite{politi1993}
and later observed also in inhibitory neural networks~\cite{Zillmer2006}.
This phenomenon is due to the prevalence of nonlinear instabilities over the linear (stable) 
evolution of the system. This leads in diffusively coupled systems 
to propagation of information (driven by nonlinear effects) and 
in diluted inhibitory networks to abrupt changes in the firing order of the neurons~\cite{politi2010stable}.
 
Clear evidences of Stable Chaos have been reported in inhibitory $\delta$--coupled
networks by considering conductance based models~\cite{Zillmer2006} as well as
current based models with time delay~\cite{JhankeTimme2008PRL,Zillmer2009LongTrans,Timme2009ChaosBalance,Luccioli2010Irregular}.
In particular, these analysis focused on the characterization of the time needed
for the transient irregular dynamics to relax to the final stable state, the
authors convincingly show that these transients diverge exponentially with the
system size, a key feature of Stable Chaos. Furthermore, in ~\cite{Zillmer2009LongTrans,Timme2009ChaosBalance}
it has been shown that, by considering time extended post-synaptic pulses
leads to a transition from stable to regular chaos, where fluctuation
driven dynamics is apparently maintained~\cite{Timme2009ChaosBalance}.

In this paper, we would like to compare the dynamics of a balanced network,
whose dynamics is driven by fluctuations in the synaptic inputs, with 
neural networks composed of tonically firing neurons.
Similar comparisons have been performed in several previous studies~\cite{tiesinga2000,renart2007}, 
however here we would like to focus on the role of nonlinear instabilities and
in particular on indicators capable to measure finite amplitude instabilities in such networks.
The effect of finite perturbations is relevant from the point of view of 
neuroscience, where the analysis is usually performed at the level of spike trains,
and a {\it minimal} perturbation corresponds to the removal or addition 
of a spike. This kind of perturbations can produce a detectable modification
of the firing rate {\it in vivo} in the rat barrel cortex~\cite{London2010}.
This has been reported as the first experimental demonstration of the sensitivity 
of an intact network to perturbations {\it in vivo}, or equivalently of an 
erratic behavior in neural circuits. It is however unclear if this sensitivity should 
be associated to linear or nonlinear effects. In particular the authors 
in~\cite{London2010} considered a network composed of excitatory and inhibitory neurons,
where an extra spike in the excitatory network is soon compensated
by an extra spike in the inhibitory network, indicating
a sort of balance in the activity of the studied neural
circuit. The ability of a perturbed balanced network
to restore rapidly the steady firing rate has been discussed
also in~\cite{MonteforteBalanced2012} for a minimal model.
Furthermore, Zillmer {\it et al.}~\cite{Zillmer2009LongTrans} 
have shown that a finite perturbation in a stable regime can cause a 
divergence of the trajectories.
These further studies, together with the fact that the addition
of an extra spike is clearly a finite perturbation from the point of view
of dynamical systems, suggest that the results reported in~\cite{London2010} 
can represent an experimental verification of Stable Chaos.

Even though all these findings are congruent with the nature of Stable Chaos
\cite{politi2010stable}, a careful characterization of this regime
in neural networks in terms of finite amplitude indicators is still lacking.
The only previous study examining this aspect in some details 
concerns a purely inhibitory recurrent Leaky Integrate-and-Fire (LIF) neural 
network with an external excitatory drive, which can sustain balanced 
activity~\cite{MonteforteBalanced2012}. Starting from this analysis,
which was limited to $\delta$-pulses, we have considered an extension 
the model to finite width pulses. Furthermore, we have characterized
the linearized evolution via usual Lyapunov exponents and
the nonlinear effects in terms of the response of the system to
finite perturbations. This analysis has been performed by employing
previously introduced indicators, like  Finite Size
Lyapunov Exponents (FSLEs)~\cite{Aurell1996FSLE} or
the probability that a finite perturbation can be 
(exponentially) expanded~\cite{MonteforteBalanced2012}, and 
new indicators capable to capture nonlinear instabilities.

The paper is organized as follows: Section \ref{sec:model} is devoted
to the introduction of the neural network model used through this paper, 
together with the indicators able to characterize linear and nonlinear instabilities. 
Section \ref{sec:results} presents a comparative study of the linear and
nonlinear stability analysis with emphasis on the influence of the pulse-width and of 
the size of the network on the dynamical behavior. Finally, in Sect. \ref{sec:discussion}
we discuss our results with respect to the existing literature and we report possible 
future developments of our research.

\section{Model and methods}
\label{sec:model}

We will consider a network of $N$ Leaky Integrate-and-Fire (LIF) neurons,
where the membrane potential $v_i$ of the $i$-th neuron  evolves
as
\begin{equation}
\label{eq:mebrane}
\dot{v}_{i}(t)= a-v_{i}(t)+I_i(t)\, \quad\quad i=1,\cdots, N \quad ,
\end{equation}
where $a > 1$ is the supra-threshold neuronal excitability, and $I_i$ represents the
synaptic current due to the pre-synaptic neurons projecting on the neuron $i$. 
Whenever a cell reaches the threshold value $v_{th} =1$ a pulse is emitted instantaneously 
towards all the post-synaptic neurons, and its potential is reset to $v_r = 0$. 
The synaptic current $I_i(t)= g E_i$ is the superposition of the pre-synaptic pulses $s(t)$
received by the neuron $i$ with synaptic strength $g$, therefore the expression of the 
field $E_i$ reads as
\begin{equation}
\label{eq:explicitE}
E_i(t) = \frac{1}{K^{\gamma}} \sum_{j \ne i} \sum_{n|t_n < t} C_{ij} 
\Theta(t-t_n) s(t-t_n) \quad .
\end{equation}
Where the sum extends to all the spikes emitted in the past in the network,
$\Theta(t-t_n)$ is the Heaviside function and the parameter $\gamma$ controls the scaling 
of the  normalization factor with the number $K$ of pre-synaptic neurons. Proper normalization
ensures homeostatic synaptic inputs \cite{turrigiano1998,turrigiano2008}.
The elements of the $N \times N$ connectivity matrix $C_{ij}$  are one (zero) in presence
(absence) of a connection from the pre-synaptic $j$-th neuron to the 
post-synaptic $i$-th one. In this paper we limit our analysis to
random sparse networks, where each neuron receives exactly $K$
pre-synaptic connections and this number remains fixed for any system size $N$.
The model appearing in Eqs. (\ref{eq:mebrane}) and (\ref{eq:explicitE}) is
adimensional, the transformation to physical units is discussed 
in Appendix I.

By following \cite{Abbott1993Asynchronous}, we assume that the pulses are
$\alpha$-functions, $s(t)=\alpha^2 t \exp(-\alpha t)$, in this case
the dynamical evolution of the fields $E_i(t)$ is ruled by the following 
second order differential equation (ODE):
\begin{equation}
\label{eq:E_secondOrder}
\ddot E_i(t) +2\alpha\dot E_i(t)+\alpha^2 E_i(t)= 
\frac{\alpha^2}{K^\gamma}  \sum_{j \ne i} \sum_{n|t_n < t} C_{ij} \delta(t-t_n) \qquad ,
\end{equation}
which can be conveniently rewritten as two first ODEs, as
\begin{equation}
\label{eq:E_and_P}
\dot{E_i} = P_i - \alpha E_i, \qquad \dot{P_i}=-\alpha P_i + 
\frac{\alpha^2}{K^\gamma}  \sum_{j \ne i} \sum_{n|t_n < t} C_{ij} \delta(t-t_n) \ ;
\end{equation}
by introducing the auxiliary field $P_i = \dot E_i -\alpha E$.

The equations \eqref{eq:mebrane} and \eqref{eq:E_and_P} can be exactly
integrated from the time $t=t_n$, just after the deliver of the
$n$-th pulse, to time $t=t_{n+1}$ corresponding to the emission
of the $(n+1)$-th spike, thus obtaining an {\it event driven map}
\cite{Zillmer2007,Olmi2010Oscillations} which reads as
\begin{subequations}
\label{eq:map}
\begin{align}
\label{eq:E_map}
E_i(n+1)&=E_i(n) {\rm e}^{-\alpha \tau(n)}+P_i(n)\tau(n) 
{\rm e}^{-\alpha \tau(n)} \\
\label{qq}
P_i(n+1)&=P_i(n)e^{-\alpha \tau(n)}+C_{im} \frac{\alpha^2}{K^\gamma}
\\
\label{V_map}
v_{i}(n+1)&=v_i(n)e^{-\tau(n)}+a(1-e^{-\tau(n)})+g H_i(n) \, ,
\end{align}
\end{subequations}
where $\tau(n)= t_{n+1}-t_n$ is the inter-spike interval associated
to two successive neuronal firing in the network, 
which can be determined by solving the transcendental equation
\begin{equation}
\label{eq:tau_implicit}
\tau(n)=\ln\left[\frac{a-v_m(n)}{a+g H_m(n)-1}\right] \ ,
\end{equation}
here $m$ identifies the neuron which will fire at time $t_{n+1}$ by reaching the
threshold value $v_m(n+1) = 1$.

The explicit expression for $H_i(n)$ appearing in equations (\ref{V_map}) and (\ref{eq:tau_implicit}) is
\begin{eqnarray}
\label{eq:H_i}
H_i(n) &=& \frac{{\rm e}^{-\tau(n)} - e^{-\alpha\tau(n)}}{\alpha-1}
     \left(E_i(n)+\frac{P_i(n)}{\alpha-1} \right)-\frac{\tau(n) e^{-\alpha\tau(n)}}{\alpha-1} P_i(n) \, .
\end{eqnarray}

The model is now rewritten as a discrete-time map with $3 N -1$ degrees
of freedom, since one degree of freedom, $v_m(n+1) =1$, is lost due
to the event driven procedure, which corresponds to 
perform a Poincar\'e section any time a neuron fires.

Our analysis will be devoted to the study of sparse networks,
by considering a constant number $K$ of afferent synapses
for each neuron, namely $K=20$.
Therefore, the normalization factor $K^\gamma$
appearing in the definition of the pulse amplitude is somehow
irrelevant, since here we limit to study a specific value of the
in-degree connectivity, without varying $K$. However, to compare with previous studies,
we set $\gamma=1$ for purely excitatory neurons, where $g>0$,
similarly to what done in \cite{Tattini2011Coherent,Luccioli2012PRL},
and $\gamma=1/2$ for purely inhibitory networks, where $g < 0$,
following the normalization employed in \cite{JhankeTimme2008PRL,Timme2009ChaosBalance,monteforte2010dynamical,MonteforteBalanced2012}.
The reasons for these different scalings rely on the fact
that in the excitatory case, the dynamics of the system are {\it current
driven} (i.e. all neurons are tonically firing even in absence
of coupling, being supra-threshold), therefore the synaptic input
should be normalized with the number of afferent neurons to maintain
an average homeostatic synaptic input~\cite{turrigiano1998,turrigiano2008}.
The situation is different in presence of inhibitory coupling, here
the supra-threshold excitability of the single neuron can be
{\it balanced} by the inhibitory synaptic currents, thus
maintaining the neurons in proximity of the firing threshold.
In this case, the network dynamics is {\it fluctuation driven}, 
because the fluctuations in the synaptic inputs  are responsible
of the neuronal firing. In order to keep the amplitude
of the fluctuations of the synaptic current constant, the normalization is
now assumed proportional to the square root of the number of the
synaptic inputs~\cite{NeuronalNetwroks2005Vogels}.
In the present analysis we have tuned the model parameters in order to be in a
fluctuation driven regime whenever the inhibitory coupling
is considered. In particular, we will study, not only the dependence of the
dynamics on the pulse shape, but also on the system size, 
by maintaining a constant number of incoming connections $K$. 
However, we will not assume that the excitatory external drive (in our
case represented by the neuronal excitability $a$) will diverge proportionally to $\sqrt{K}$, 
as done in~\cite{vanVreeswijk1996,MonteforteBalanced2012}, since
we are not interested in the emergence of a {\it self-tuned} balanced state in the limit
$ K \to \infty$, for $1 << K << N$~\cite{vanVreeswijk1996,MonteforteBalanced2012}.

\subsection{Linear Stability Analysis}
\label{sec:lyapunov}

To perform the linear stability analysis of the system, we
follow the evolution of an infinitesimal perturbation in the
tangent space, through the following set of
equations obtained from the linearization
of the event driven map (\ref{eq:E_map},\ref{qq},\ref{V_map})
\begin{subequations}
\label{eq:lin1}
\begin{align}
\delta E_i(n+1) &=  e^{-\alpha \tau(n)} \left[ \delta E_i(n) +\tau(n) \delta
P_i(n) \right] \nonumber
\\
 & - e^{-\alpha \tau(n)} \left[\alpha E_i(n) + (\alpha \tau(n)-1) P_i(n) \right] \delta \tau(n)\,,\\
\delta P_i(n+1) &=  e^{-\alpha \tau(n)} \left[ \delta P_i(n)-\alpha P_i(n) \delta \tau(n) \right]\, ,
\\
\delta v_{i}(n+1) &=  e^{-\tau(n)} \left[ \delta v_i (n) + (a-v_i(n)) \delta \tau(n) \right] + g \delta H_i (n) 
\nonumber
\\
i & =1,\dots,N \quad ; \quad \delta v_m(n+1) \equiv 0 \, .
\end{align}
\end{subequations}
The boundary condition $\delta v_m(n+1) \equiv 0$ is a consequence of the event driven
evolution. The expression of $\delta \tau(n)$ can be computed by differentiating 
\eqref{eq:tau_implicit} and \eqref{eq:H_i}
\begin{eqnarray}
\delta \tau(n) =\tau_v \delta v_m(n) +\tau_E\delta E_m(n)+\tau_P\delta P_m(n)\ ,
\end{eqnarray}
where
\begin{eqnarray}
\tau_v:= \frac{\partial \tau}{\partial v_m}  \quad  , \quad
\tau_E:= \frac{\partial \tau}{\partial E_m}  \quad ,  \quad
\tau_P:= \frac{\partial \tau}{\partial P_m}  \quad .
\end{eqnarray}

In this paper, we will limit to measure the maximal Lyapunov exponent $\lambda$
to characterize the linear stability of the studied models. This
is defined as the the average growth rate of the infinitesimal perturbation 
$$ \bm{\delta} = (\delta v_1 \dots \delta v_N,\, \delta E_1 \dots \delta E_N, \, \delta P_1 \dots \delta P_N),$$ 
through the equation 
\begin{equation}
\label{eq:def_lyapunov}
\lambda=\lim_{t\to\infty}\frac{1}{t}\log\frac{\mid \bm{\delta} (t)\mid}{\mid \bm{\delta}_0\mid} \quad ,
\end{equation}
where $\bm{\delta}_0$ is the initial perturbation at time zero. The
evolution of the perturbation $\bm{\delta}(t)$ has been followed
by performing at regular time intervals the rescaling of its
amplitude to avoid numerical artifacts, as detailed in \cite{BenettinLyapunov1980}.
Furthermore, since our system is time continuous one would expect to have always a zero
Lyapunov exponent, which in fact is the maximal Lyapunov if the system is 
not chaotic. However, this does not apply to the event driven map because
the evolution is based on a discrete time dynamics, where the motion
along the orbit between two successive spikes is no more present due to the performed
Poincar\'e section.

\subsection{Finite Size Stability Analysis}

  Besides the characterization of the stability of infinitesimal perturbations,
we are also interested in analyzing how a perturbation grows according to its amplitude.
To perform this task several indicators have been introduced in the last years,
ranging from {\it Finite Size Lyapunov Exponents} (FSLE)~\cite{Aurell1996FSLE,LetzKantz2000,Cencini2010chaos,FSPolitiLE}
to the propagation velocity of finite perturbations~\cite{Torcini1995FSLE}. 
FSLEs have been mainly employed to charaterize {\it Stable Chaos} 
in spatially extended systems \cite{politi2010stable} and Collective
Chaos in globally coupled systems \cite{shibata1998, cencini1999,Olmi2010Chaos}. 

We have performed several tests by employing the usual FSLE definition \cite{FSPolitiLE}. In particular FSLE can be defined by considering an unperturbed
trajectory $\bm{x} = (v_1 \dots v_N,\, E_1 \dots  E_N, \,  P_1 \dots  P_N)$
and a perturbed trajectory $\bm{x^\prime} = (v_1^\prime \dots v_N^\prime,\, E_1^\prime \dots  E_N^\prime, \,  P_1^\prime \dots  P_N^\prime)$, obtained by randomly perturbing all the coordinates (both the fields $E$ and $P$ as well as the membrane potentials) 
of the generic configuration $\bm{x}$ on the attractor. Then
we follow the two trajectories in time by measuring their distance
$\Delta(t)= \parallel \bm{x}(t) - \bm{x'}(t) \parallel$, by employing the absolute value norm.
Whenever $\Delta(t_k)$ crosses (for the first time) a series of exponentially spaced
thresholds $\theta_k$, where $\theta_k = r \theta_{k-1}$, the crossing times $t_k$ are
registered. By averaging the time separation between consecutive crossings over different 
pairs of trajectories, one obtains the FSLE \cite{FSPolitiLE, Aurell1996FSLE}
\begin{equation}
\lambda_F (\Delta(t_k)) = \frac{r}{\langle t_k - t_{k-1} \rangle}
\qquad ; {\rm where} \quad \Delta(t_k)=\theta_k
\label{FSLE}
\end{equation}
For small enough thresholds, one recovers the usual maximal Lyapunov exponent, 
while for large amplitudes, FSLE saturates to zero, since a perturbation cannot be larger 
than the size of the accessible phase-space. In the intermediate range, $\lambda_F$ tells 
us how the growth of a perturbation is affected by nonlinearities. 
However, as a general remark, 
we have noticed that it is extremely difficult to get reliable results
from the FSLE analysis, probably because the estimation of $\lambda_F$ relies on measurements 
based on single trajectory realizations, which presents huge fluctuations. 
In order to overcome this problem, the single trajectory should be smoothed before
estimating the passage times from one threshold to the next one and we observed 
that the results strongly depend on the adopted smoothing procedure, 
in particular for the fluctuation driven case. 

Therefore, in order to investigate the growth rate of finite amplitude perturbations we have decided
to adopt different indicators rather than the FSLE. In particular,
an estimation of finite size stability can be obtained by defining the following indicator
\begin{equation}
\label{indicator}
D(\Delta(t)) = \frac{ d \left\langle \log  \Delta (t)  \right\rangle}{dt}
\enskip ;
\end{equation}
where the average $\left\langle \cdot \right\rangle$ is performed over many different
initial conditions. In the limit $\Delta (t) \to 0$ we expect to recover the maximal
Lyapunov exponent $\lambda$. In order to ensure that the dynamics of the perturbed
trajectory will also occur on the attractor associated to the studied dynamics, 
we have considered extremely small initial perturbations 
$\Delta_0 = \Delta (0) \simeq 10^{-8}-10^{-10}$. As we will show, after a transient
needed for the perturbed trajectory $\bm{x^\prime}$ to relax to the attractor, 
$D(\Delta)$ measures effectively the maximal Lyapunov exponent.
However, if nonlinear mechanisms are present $D(\Delta)$ can become larger
than $\lambda$ for finite amplitude perturbations. Anyway, analogously to the
FSLE, for perturbations of the size of the attractor the indicator 
$D(\Delta)$ decays towards zero due to the trajectory folding.

The studied models present discontinuities of ${\cal O}(1)$ in the membrane potentials $v_i$, due
to the reset mechanisms, and of ${\cal O}(\alpha^2/K^\gamma)$ in the fields $P_i$, due to the pulse 
arrival. In order to reveal, without any ambiguity, the presence of nonlinear instabilities
at finite amplitudes, for the estimation of the FSLE and of the indicator $D$
we mainly limit our analysis to the continuous fields $\{ E_i \}$.
In particular, to characterize the finite amplitude instabilities, we consider 
the following distance between the perturbed and unperturbed orbits
\begin{equation}
\label{perturb_E}
\Delta^{(E)}(t) = \frac{1}{N} \sum_{i=1}^N | E_i(t) - E_i^\prime (t)|
\enskip .
\end{equation}
In some cases we have also analyzed the distance $\Delta^{(v,E,P)}$ between all the variables
associated to the unperturbed and perturbed state with a clear meaning of the adopted symbol.

Unfortunately, the indicator $D(\Delta)$ as well as the FSLE cannot be employed
in the case of stable chaos, when $\lambda < 0$, because in this case small perturbations
are quickly damped and one cannot explore the effect of perturbation of growing
amplitude by following the dynamics on the attractor. In this situation, one 
should employ different indicators, as done in \cite{LetzKantz2000,CenciniTorcini}
for coupled map lattices. In particular, we proceed as follow, we consider
two orbits at an initial distance $\Delta_0$ and we follow them for a time interval $T$,
then we measure the amplitude of the perturbation  at the final time, namely $\Delta (T)$.
We rescale one of the two orbits to a distance $\Delta_0$ from the other one, keeping the 
direction of the perturbation unchanged, and we repeat the procedure several times
and for several values of $\Delta_0$. Then, we estimate the finite amplitude growth rate, as
\begin{equation}
R_T (\Delta_0) = \frac{1}{T}\left\langle \log \frac{ | \Delta (T) | }{|\Delta_0|} \right\rangle ,
\end{equation}
where the angular brackets denote the average over a sufficiently large number of repetitions.
To allow the perturbed orbit to relax on the attractor, we initially perform 
$\simeq 10^3$ rescalings, which are not included in the final average. However, also this
procedure does not guarantee that the attractor is always reached, in particular for
very large perturbations. Furthermore, the perturbed dynamics is no more constrained to 
evolve along the tangent space associated to the event driven map. As a matter of fact, whenever 
$\lambda < 0$ the indicator $R_T(\Delta_0)$ converges to zero 
and not to the Lyapunov exponent associated to the discrete time map evolution. 
 
Finally, following the analysis reported in~\cite{MonteforteBalanced2012},
we consider the probability $P_S(\Delta_0)$ that a perturbation of amplitude $\Delta_0$ induces 
an exponential separation between the reference and perturbed trajectory. In particular, we perturb 
the reference orbit with an initial perturbation $\Delta_0$ and we follow the evolution
of the trajectories for a time span $T$. Whenever $\Delta(T)$ is larger than a certain threshold
$\theta_L$ this trial contributes to the number of expanding initial perturbations 
$N_S(\Delta_0)$, otherwise is not counted. 
We repeat this procedure $N_T$ times for each perturbation of amplitude $\Delta_0$,
then $P_S(\Delta_0)= N_S(\Delta_0) /N_T$.
For the two latter indicators, namely $R_T$ and $P_S$, we have always employed the total distance
$\Delta^{(v,E,P)}$, to confront our findings with the results reported in~\cite{MonteforteBalanced2012}.

\section{Results}
\label{sec:results}

\begin{figure}
\centering
\subfigure{\label{fig:raster_insieme}\includegraphics[width=0.48\textwidth]{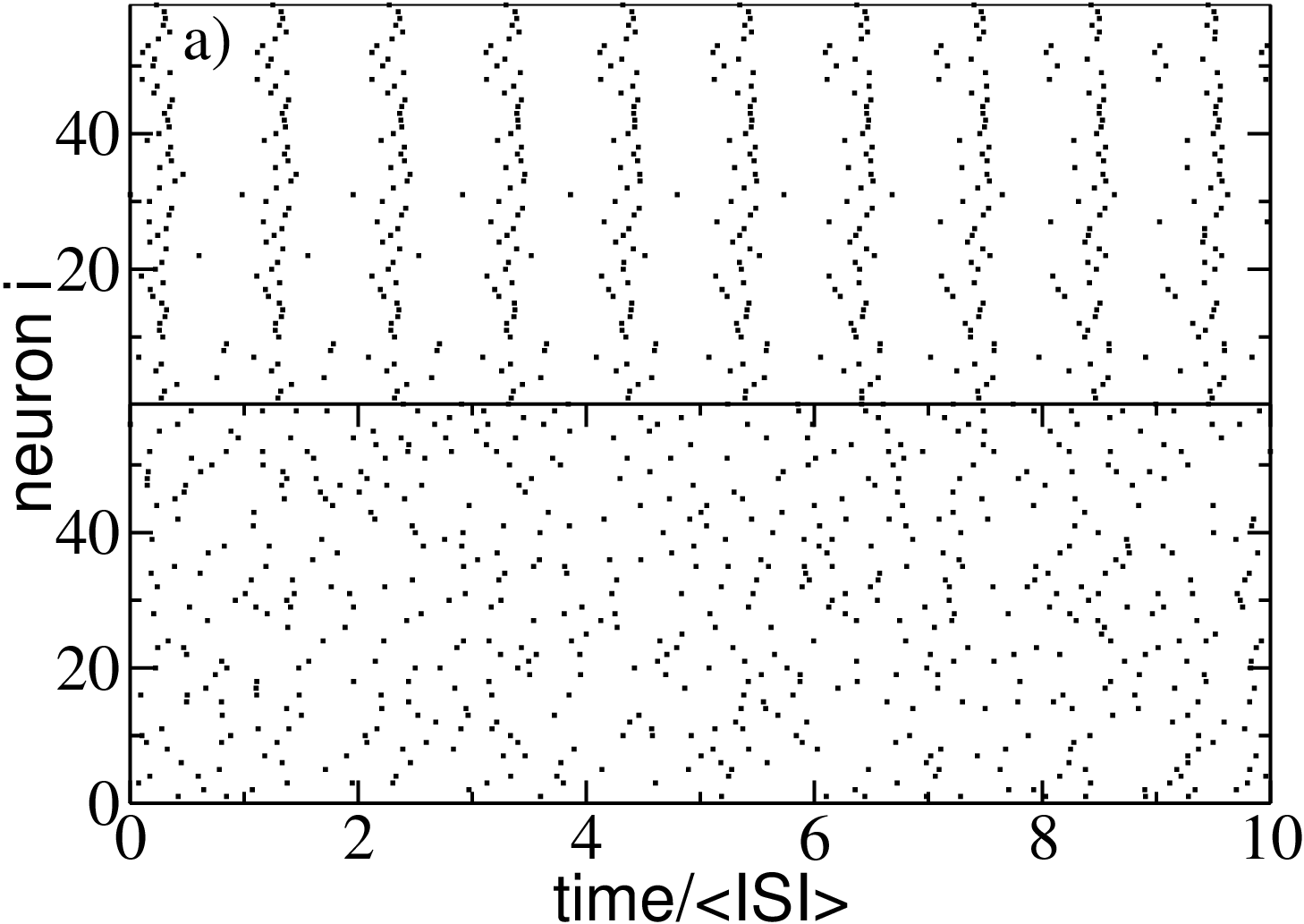}}
\subfigure{\label{fig:Vtrace_insieme}\includegraphics[width=0.48\textwidth ]{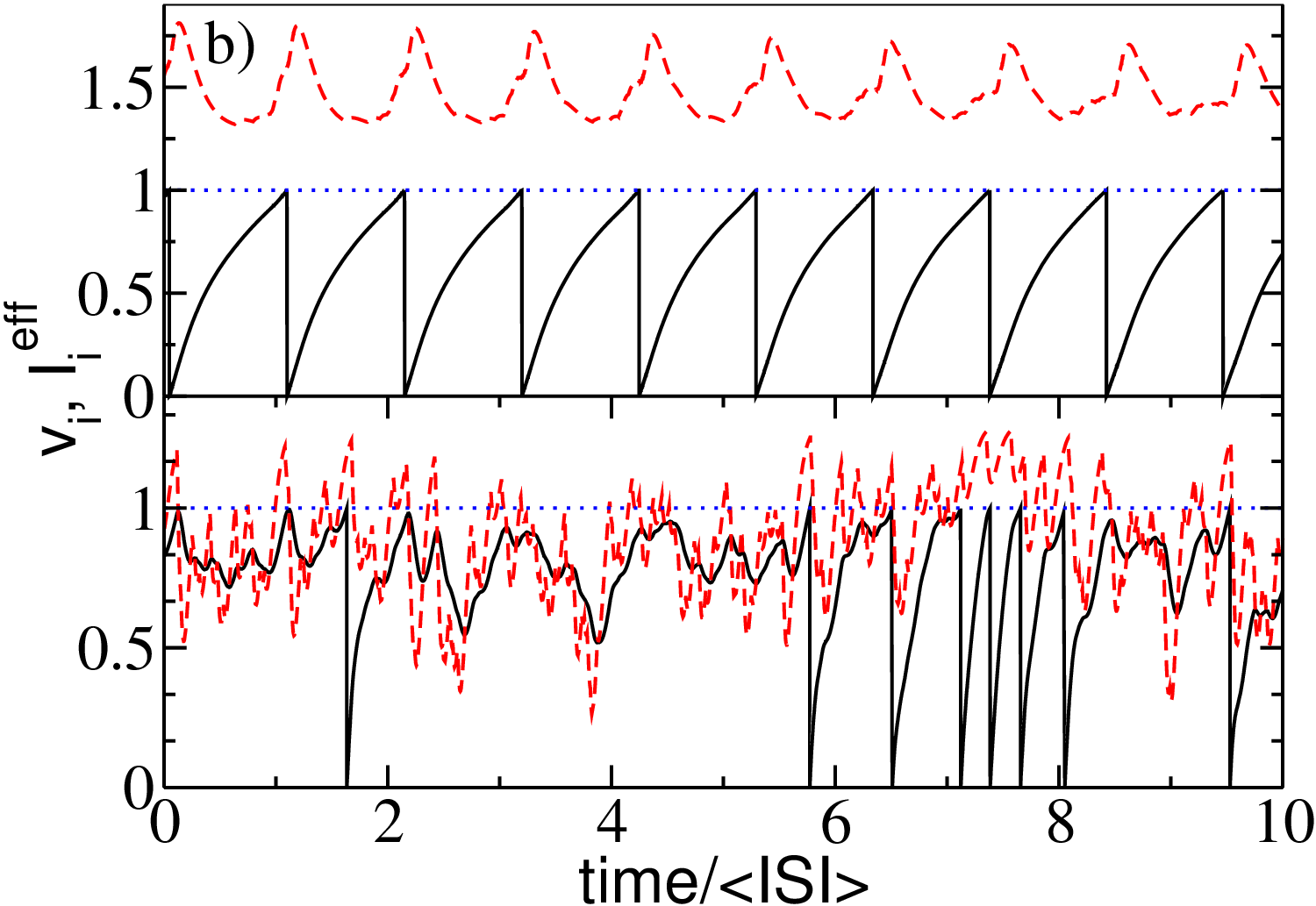}}
\caption{Comparison between current driven (upper panels) and fluctuation driven (lower panels) activity. (a) Raster
plots for a pool of $60$ neurons. (b) Membrane potential traces $v_i(t)$ (black solid line) 
and the corresponding effective current $I_i^{eff}$ (red dashed line) for a typical neuron.
The blue dotted line indicates the firing threshold.
For the current driven case, $a=1.3$, $g = 0.2$, $\alpha = 9$ and $\gamma = 1$, corresponding
to the situation studied in \cite{Luccioli2012PRL}; for the fluctuation driven network
the parameters are the same, apart $g = -0.8$, $\alpha=5$ and $\gamma = 1/2$.
For both networks, $K = 20$ and $N=400$ and the results for both systems are reported
for the same {\it rescaled} time intervals $t/<ISI> = 10$, after discarding a transient of $10^4$ spikes.}
\label{fig:rasters}
\end{figure}

As already mentioned, we will compare a current driven excitatory network
and a fluctuation driven inhibitory network. In
particular, the excitatory network is studied in a regime where
it presents a collective non trivial {\it partial synchronization}
\cite{vanVreeswijk1996Partial,Luccioli2012PRL}. This state is characterized 
by quasi-synchronous firing events, 
as revealed by the raster plot reported in the upper panel of Fig.~\ref{fig:raster_insieme},
and almost periodic oscillations of the effective current $I_i^{eff}(t) \equiv a + g E_i(t)$
(see Fig.~\ref{fig:Vtrace_insieme}, upper panel). In this particular case $I_i^{eff} > 1$ 
therefore the neurons are always supra-threshold. In this situation the measure of the $CV$ gives 
quite low values, namely for the studied case 
(with $a=1.3$, $g=0.2$ and $\alpha =9$) $CV \simeq 0.17$, 
similar to pyramidal neurons in {\it vitro}. Despite this low level of variability in 
the neuronal dynamics, the sparseness in the matrix connectivity induces chaotic 
dynamics in the network, which persists even in the thermodynamic limit~\cite{Luccioli2012PRL}.
At variance with diluted networks, where the average connectivity scales proportionally to the system
size ($K \propto N^z$, with $1 \ge z > 0$). In this latter case, in the limit $N \to \infty$
the system will recover a regular evolution, 
similarly to fully coupled networks~\cite{Olmi2010Oscillations, Tattini2011Coherent}.

\begin{figure}
\centering
\subfigure{\label{fig:cv_vs_alpha}\includegraphics[width=0.45\textwidth ]{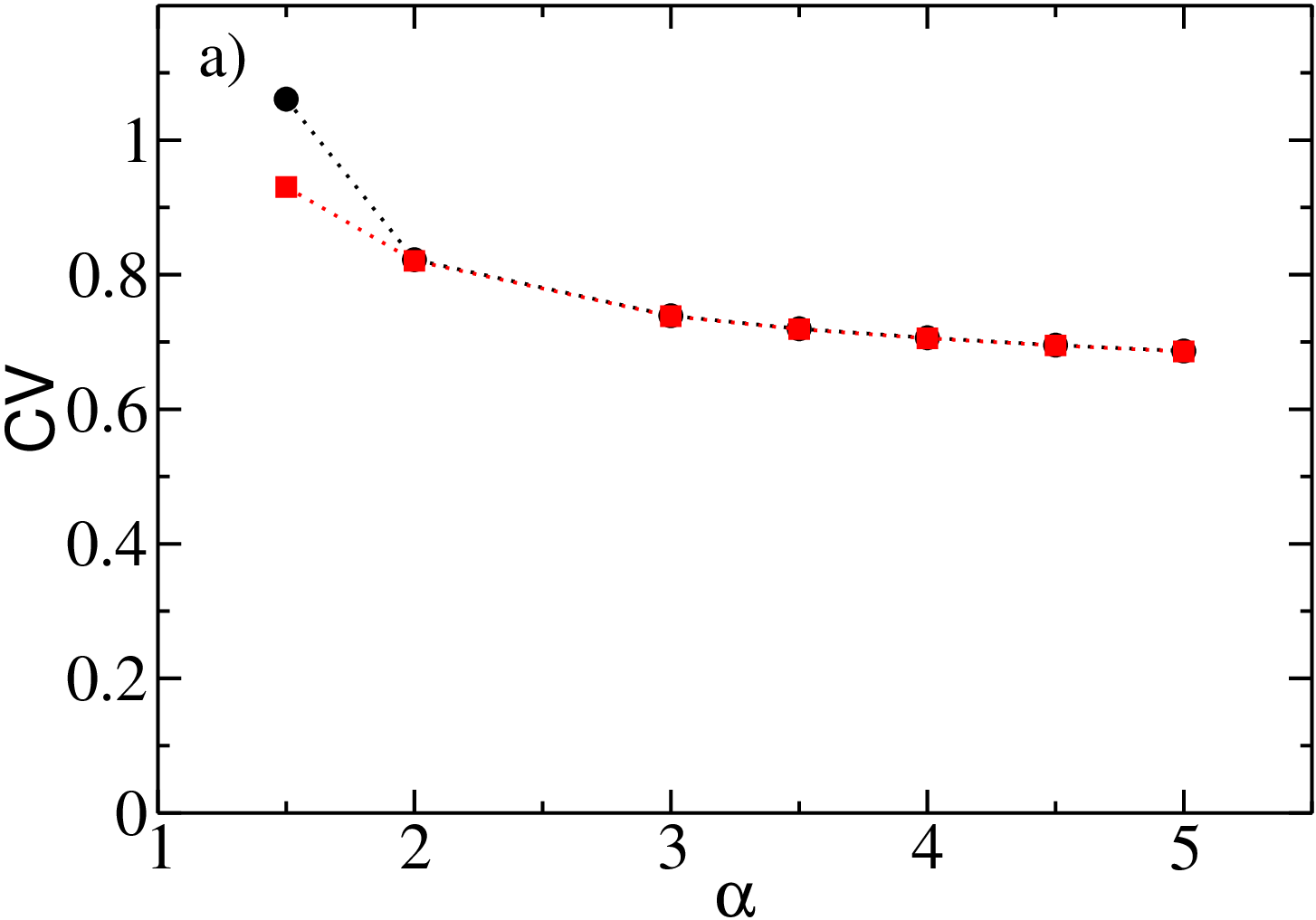}}
\subfigure{\label{fig:isi_vs_alpha}\includegraphics[width=0.45\textwidth ]{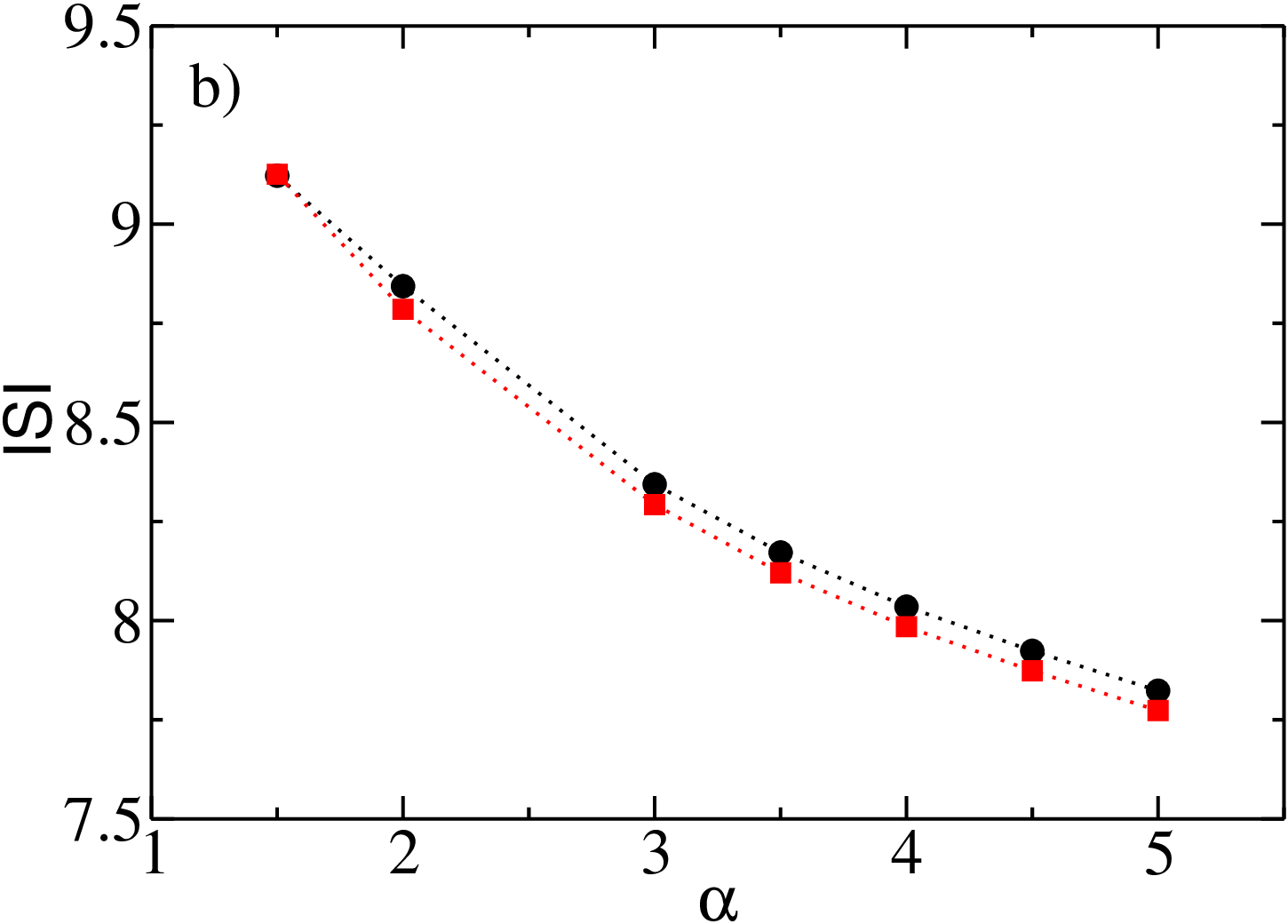}}
\caption{Dependence of the coefficient of variation $CV$ (a) and of the inter-spike time interval 
ISI (b) on the pulse width for the fluctuation driven case. The data refer to
$N = 400$ (black circles) and $N=1600$ (red squares). 
The data have been averaged over $10^8$ spikes, once a transient of $10^7$ spikes has been discarder. 
The other parameters are as in the caption of Fig. \ref{fig:rasters}}
\label{fig:CV_ISISvsalpha}
\end{figure}

\begin{figure}[htb]
\subfigure{\label{fig:cv_vs_N}\includegraphics[width=0.45\textwidth ]{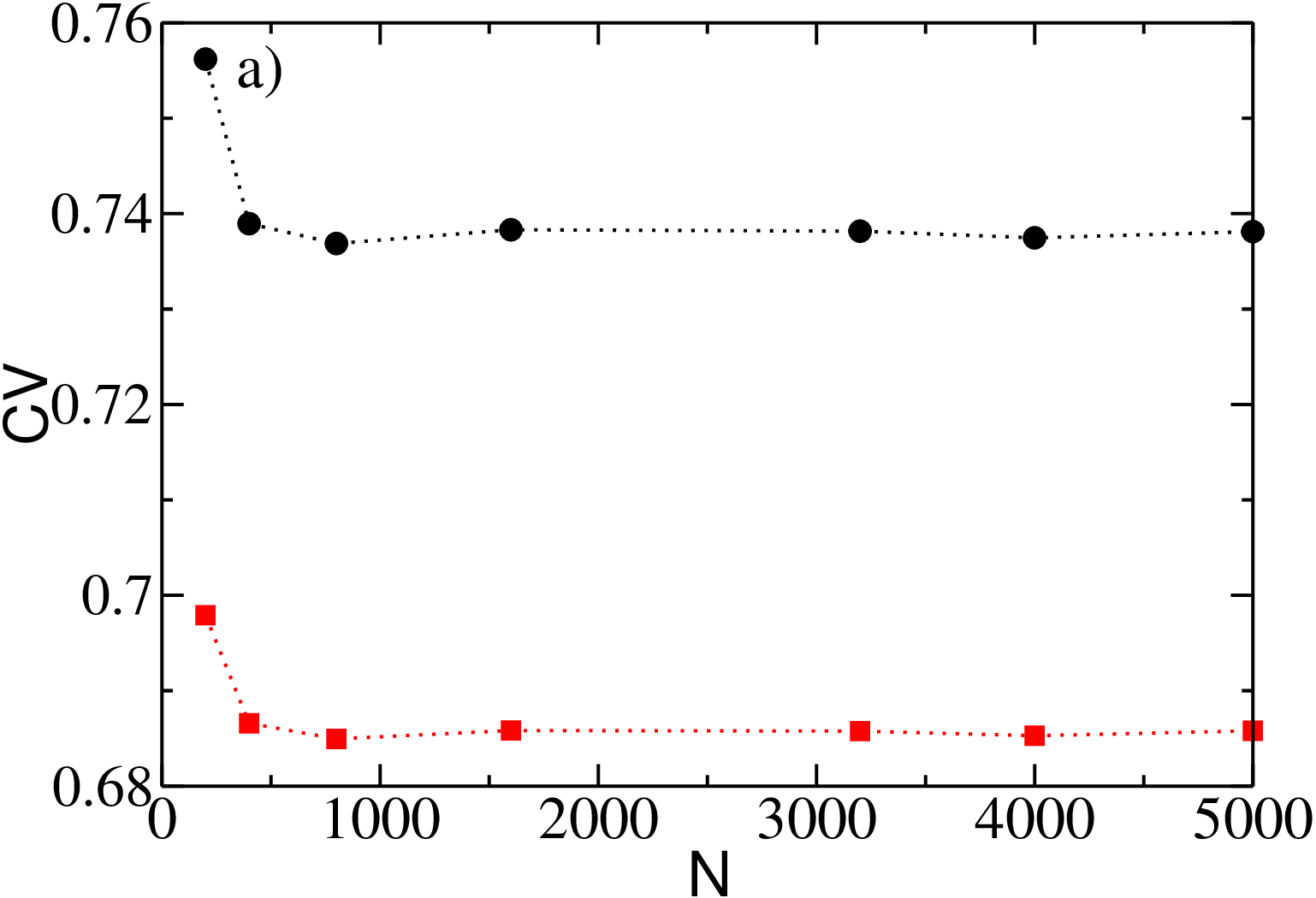}}
\subfigure{\label{fig:isi_vs_N}\includegraphics[width=0.45\textwidth ]{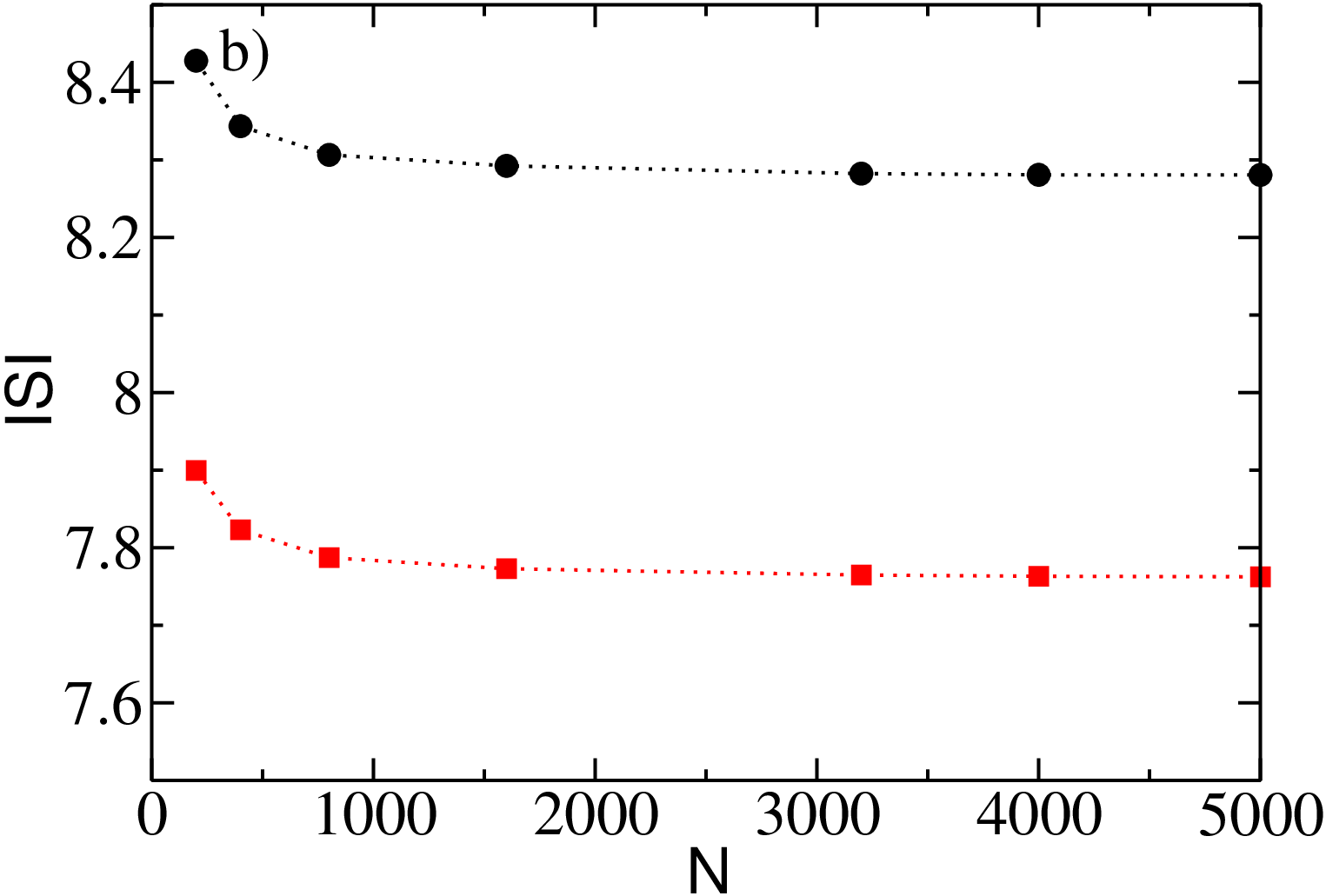}}
\caption{Dependence of the coefficient of variation $CV$ (a) and of the inter-spike time interval 
ISI (b) on the size of the network for fluctuation driven networks in two
representative situations corresponding to the chaotic ($\alpha=3$, black circles) and 
the stable chaos ($\alpha = 5$, red squares) regimes. The reported data have been averaged
over $10^8$ spikes, once a transient of $10^7$ spikes has been discarded. 
The other parameters are as in the caption of Fig. \ref{fig:rasters}.}\label{fig:CV_ISISvsSize}
\end{figure}

For the inhibitory network, we observe radically different dynamics,
this because now $I^{eff}(t)$ oscillates around one, therefore the neurons
fire in a quite irregular manner, driven by the fluctuations of the fields
$E_i(t)$, as shown in the lower panels of Figs.~\ref{fig:rasters} (a) and (b). In this case
we have examined the dynamics of the model for $a=1.3$, $g=-0.8$ and different pulse-widths $1/\alpha$.
For $\alpha \in [1:5]$ the neuronal dynamics are always quite erratic,
being characterized by $CV \simeq 0.7-1$ (see Fig.~\ref{fig:cv_vs_alpha}).
Narrower pulses (larger $\alpha$ values) are associated to somehow
more regular dynamics and smaller ISI, however we have verified that
the ISI and CV saturates to some finite value in the thermodynamic limit 
(as shown in Fig.~\ref{fig:CV_ISISvsSize} (a) and (b)). This suggests that fluctuations
will not vanish for $N \to \infty$ and that the system will remain fluctuation
driven even in such a limit. Furthermore, the two $\alpha$-values examined in
Figs. \ref{fig:cv_vs_N} and \ref{fig:isi_vs_N} correspond to two
different dynamical regimes, further discussed in Sect.~\ref{sec:lyapunov_results}, 
namely, a chaotic ($\alpha=3$) and a non-chaotic ($\alpha=5$) state.

\subsection{Lyapunov analysis}
\label{sec:lyapunov_results}

\begin{figure}[htb]
\centering
\subfigure{\label{fig:lambda_vs_alpha}\includegraphics[width=0.45\textwidth ]{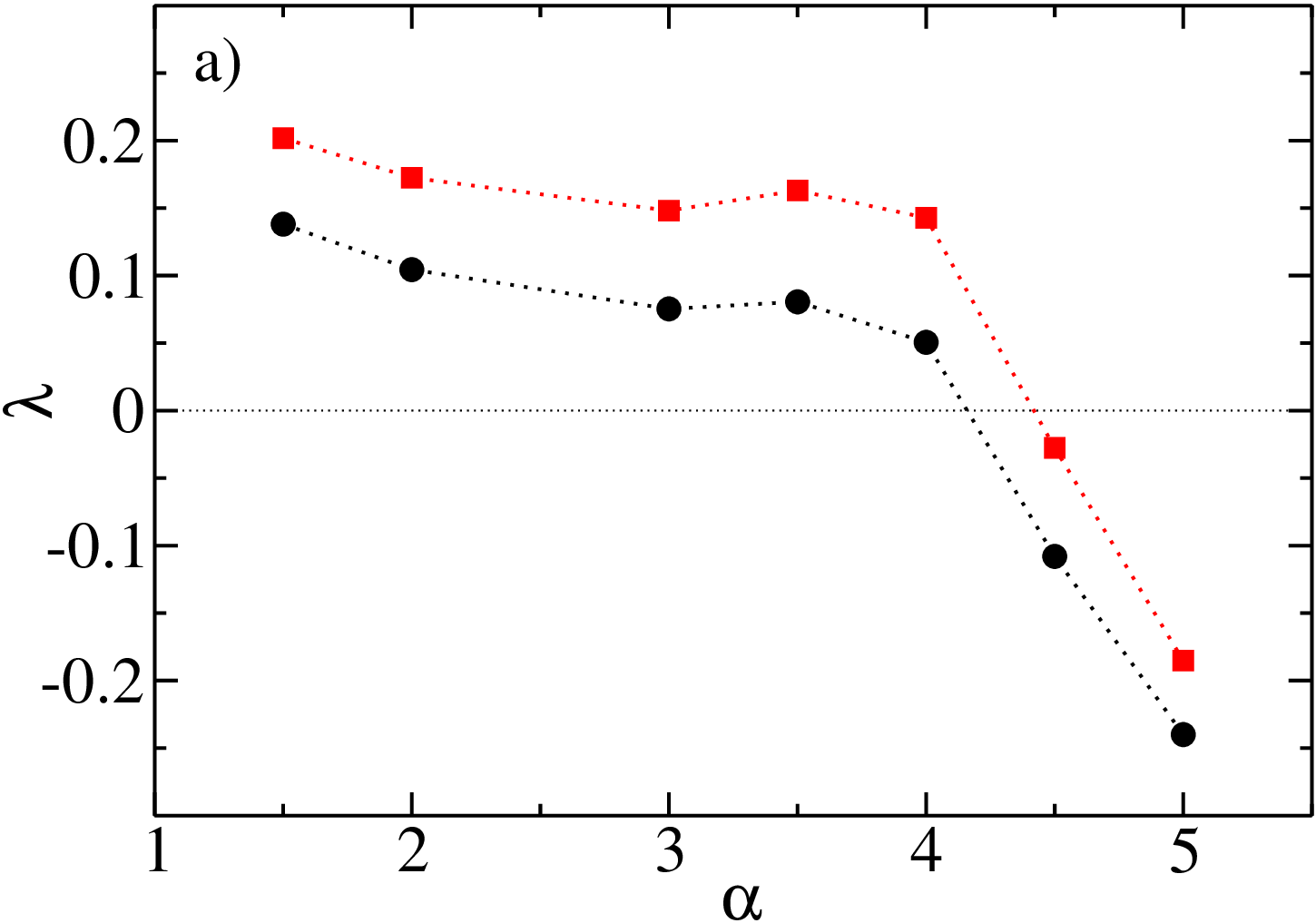}}
\subfigure{\label{fig:lambda_vs_N}\includegraphics[width=0.45\textwidth ]{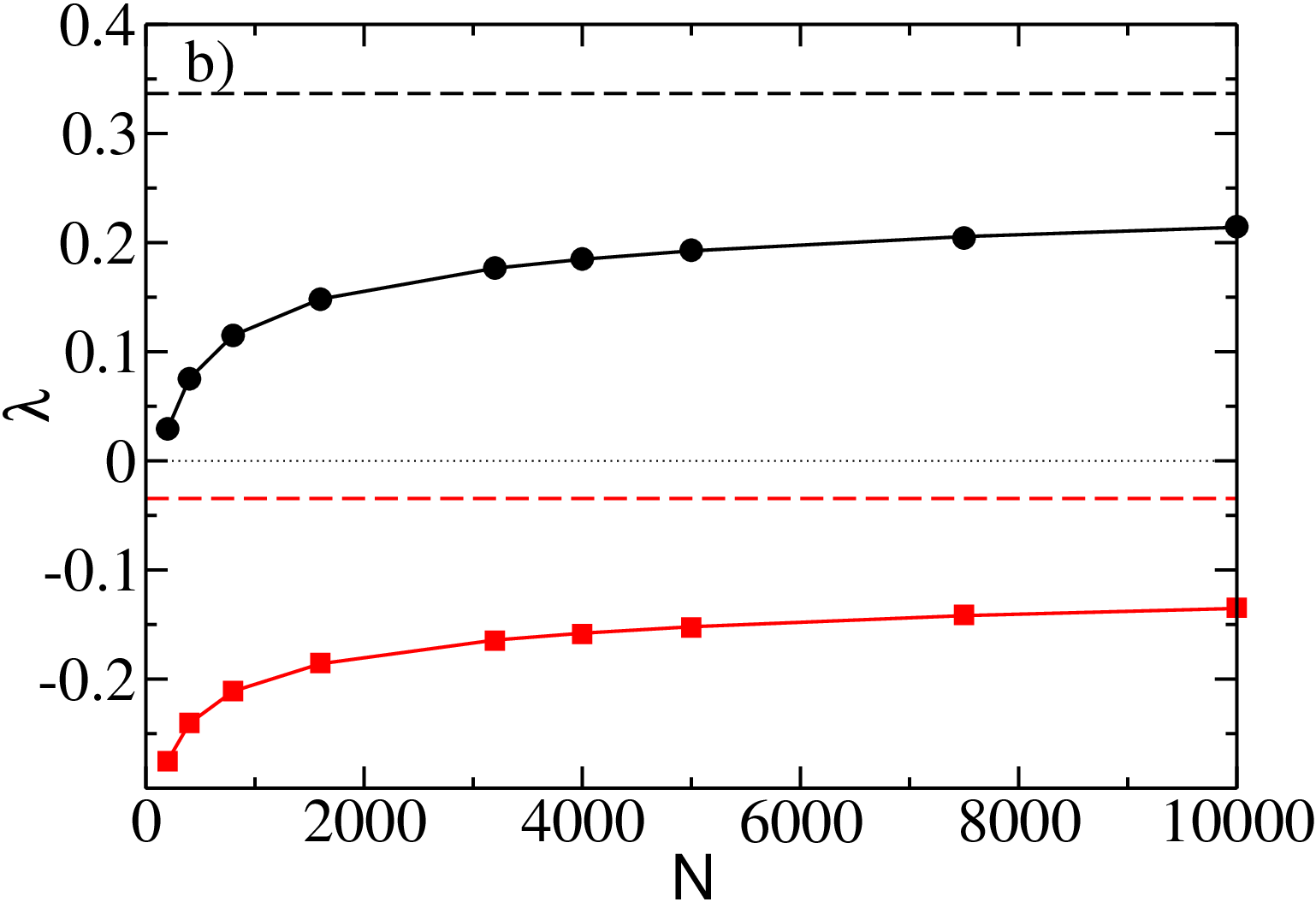}}
\caption{Linear stability analysis of the fluctuation driven state. 
(a) Maximal Lyapunov exponent $\lambda$ as a function of  pulse-width $\alpha$, 
for two representative system sizes: $N = 400$ (black circles) and 
$N = 1600$ (red squares); thin dashed lines are drawn for eye guide only.
(b) Lyapunov exponent as a function of the system size
$N$, for two representative pulse widths: $\alpha = 3$ (black circles) and $\alpha = 5$ (red squares). 
Continuous lines correspond to the nonlinear fitting \eqref{eq:fitting},
which predicts the asymptotic values $\lambda_{\infty}$ (thick dashed lines).
The fitting parameters entering in Eq.~\eqref{eq:fitting} are
$c = 1.08$ ($c = 0.78$) and $\eta \simeq 0.24$ ($\eta \simeq 0.22$)
for $\alpha = 3$ ($\alpha = 5$). In both figures, $\lambda$ is calculated by 
integrating the evolution in the tangent space together with the unperturbed orbit dynamics
during a time interval equivalent to $10^8$ spikes, after discarding a
transient of $10^7$ spikes. Remaining parameters as in Fig. \ref{fig:rasters}.}
\label{fig:lyapunov}
\end{figure}

As previously shown, the fluctuation driven regime is observable for 
the inhibitory network for all the considered pulse widths.
In this Subsection we would like to investigate whether such variability
is related to a linear instability of infinitesimal
perturbations (measured by the maximal Lyapunov
exponent $\lambda$) or to other (nonlinear) instabilities present in the system.
Let us start examining the Lyapunov exponent for such systems,
as a first result we observe a strong dependence of $\lambda$ 
on the pulse-width (see Fig.~\ref{fig:lambda_vs_alpha}): the system
is chaotic for wide pulses and becomes stable for sufficiently
narrow ones. These results are in agreement with previously reported results in~\cite{Zillmer2009LongTrans,Timme2009ChaosBalance} 
for an inhibitory network of LIF neurons with delayed synapses. In these papers
the authors show that chaos can arise only for sufficiently broad 
pulses, conversely for $\delta$-pulses the system is always stable.
It is worth to notice that the critical $\alpha$-value at which occurs the transition to
chaos becomes larger as the system size increases, pointing to the
question whether the stable regime still exists for finite pulses in 
the thermodynamic limit or if it is merely a finite size property~\cite{Timme2009ChaosBalance}.
Extensive simulations for sizes of the network up to $N=10,000$ 
have shown that the stable regime is present even for such a large size
(see Fig.\ref{fig:lambda_vs_N}). 
Furthermore, we have found an 
empirical scaling law describing the increase of $\lambda$ with
$N$, i.e.
\begin{equation}
\label{eq:fitting}
\lambda = \lambda_{\infty} - c N^{-\eta}
\end{equation}
%
%
where $\lambda_{\infty}$ denotes the asymptotic value in the thermodynamic limit
and $\eta$ is the scaling exponent. For the two representative cases here studied, 
the exponent was quite similar, namely $\eta \simeq 0.24$ ($\eta \simeq 0.22$) for $\alpha =3$ ($\alpha =5$),
thus suggesting an universal scaling law for this model when fluctuation driven,
with an exponent $\eta=1/4$. This exponent is different from the one measured
for the current driven case, in such situation for sparse connectivity $\lambda$ converged to its
asymptotic value as $1/N$~\cite{Luccioli2012PRL}. An exponent $\eta=1$ has been
previously measured for coupled map lattices exhibiting spatio-temporal chaos
and theoretically justified in the framework of the Kardar-Parisi-Zhang equation~\cite{pikovsky1998dynamic}. 
The scalings we are reporting in this paper are associated to random networks,
therefore they demand for a new theoretical analysis.
Furthermore, the asymptotic values $\lambda_{\infty}=0.335(1)$ ($\lambda_{\infty}=-0.034(1)$)
indicate that a critical threshold separating stable from chaotic dynamics
persists in the thermodynamic limit.

\subsection{Finite size perturbation analysis}

 Stable chaos in spatially extended systems is due to the propagation of finite 
amplitude perturbations, while infinitesimal ones are damped.  
In inhibitory neural networks, the origin of Stable Chaos has been ascribed  
to abrupt changes in the firing order of neurons induced by a discontinuity 
in the dynamical law, while infinitesimal perturbations leave the order unchanged~\cite{Zillmer2006,politi2010stable,Timme2009ChaosBalance}.
In particular, by examining a conductance based model, in ~\cite{politi2010stable}
it has been shown that  
a spike was able to induce a finite perturbation  in the  evolution of two (not-symmetrically) connected neurons, 
given that the inhibitory effect of a spike was related to the actual value of the membrane potential of the receiving neuron. Therefore two ingredients are needed to observe Stable Chaos in neural models, 
a non symmetric coupling among neurons, together with the fact that the amplitude
of transmitted pulses should depend on the neuron state.
These requirements are fulfilled also in the present model, despite being current based, 
since any current based model can be easily transformed in a conductance based
one via a nonlinear transformation~\cite{Abbott1993Asynchronous, politi2010lif}.
However, the problem is to quantify this effect in terms of some indicator,
similarly to what done in spatially extended systems, where Stable Chaos has been characterized 
in terms of the FSLE and of the velocity of propagation of information \cite{Torcini1995FSLE,CenciniTorcini}.

\begin{figure}[htb]
\centering
\includegraphics[width=0.6\textwidth ]{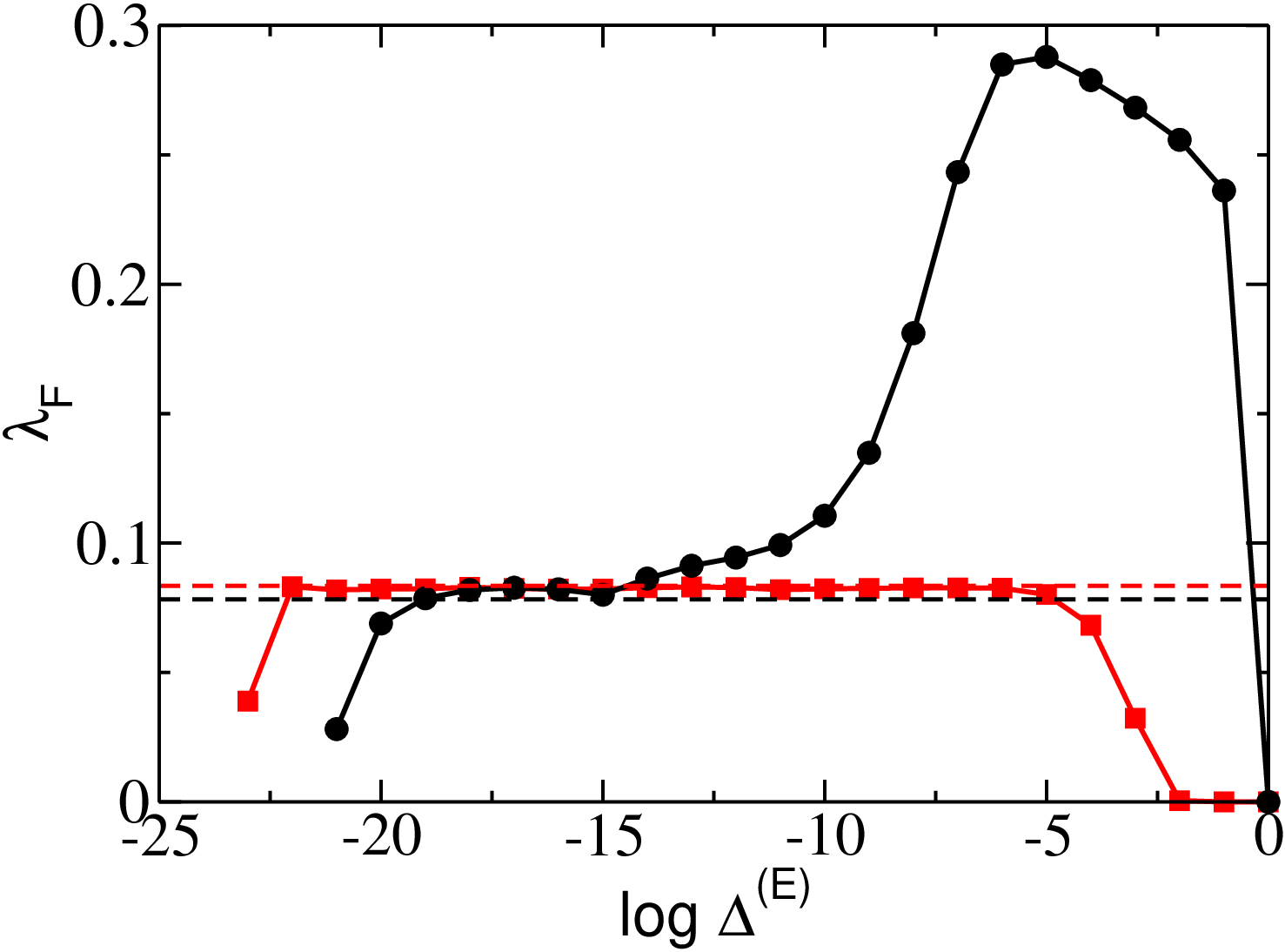}
\caption{FSLE indicator $\lambda_F$ for the fluctuation driven (black circles) and current driven (red squares) chaotic set-ups. An initial perturbation of $10^{-9}$ ($10^{-7}$) is applied to the excitatory (inhibitory) network. The distance between the perturbed and unperturbed trajectory $\Delta^{(E)}$ is sampled during 300 time units, at fixed time intervals  $dt = 0.2$. The sampled curve is smoothed over a sliding window of 20 time units and the resulting curve
is used to obtain the times $t_k$ at which the system crosses the corresponding thresholds $\theta_k$, with $r=1$
(see the definition (\ref{FSLE}). This procedure is averaged in the current (fluctuation) driven
case over 5000 (15000) realizations. 
Thick dashed lines indicate the value of $\lambda$ for each one of the two cases.
The current and fluctuation driven cases have been examined
for the same parameter values reported in Fig. \ref{fig:rasters}, apart that
for the inhibitory case the inverse of the pulse width is
set to $\alpha=3$.
}
\label{fig:FSLE}
\end{figure}

 As a first indicator we consider the FSLE, associated to the norm $\Delta^{(E)}$, 
the corresponding results are reported in Fig.~\ref{fig:FSLE} for the current and fluctuation driven
cases. In the former case the FSLE is never larger than the usual Lyapunov exponent $\lambda$,
with which it coincides over a wide range of perturbation amplitudes. In particular, 
$\lambda_F (\Delta ^{(E)}) < \lambda$ for small amplitudes, due to the fact that initially the perturbation needs 
a finite time to align along the maximal expanding direction. Furthermore, due to
the folding mechanism, the perturbation is contracted also for large perturbations of the 
order of the attractor system size. In summary, for current driven dynamics
only the instability associated to infinitesimal perturbation is present,
as reported also in~\cite{Olmi2010Chaos}. In the fluctuation driven case the situation
is quite different as shown in Fig.~\ref{fig:FSLE}, the FSLE 
essentially coincides with $\lambda$ for small $\Delta^{(E)}$, but it becomes definitely
larger than $\lambda$ for finite perturbations, revealing a peak around 
$\Delta^{(E)} \simeq {\cal O}(1/N)$. These are clear indications that finite 
amplitude instabilities coexist with infinitesimal ones and they could be in principle 
even more relevant.

\begin{figure}[htb]
\centering
\includegraphics[width=0.6\textwidth ]{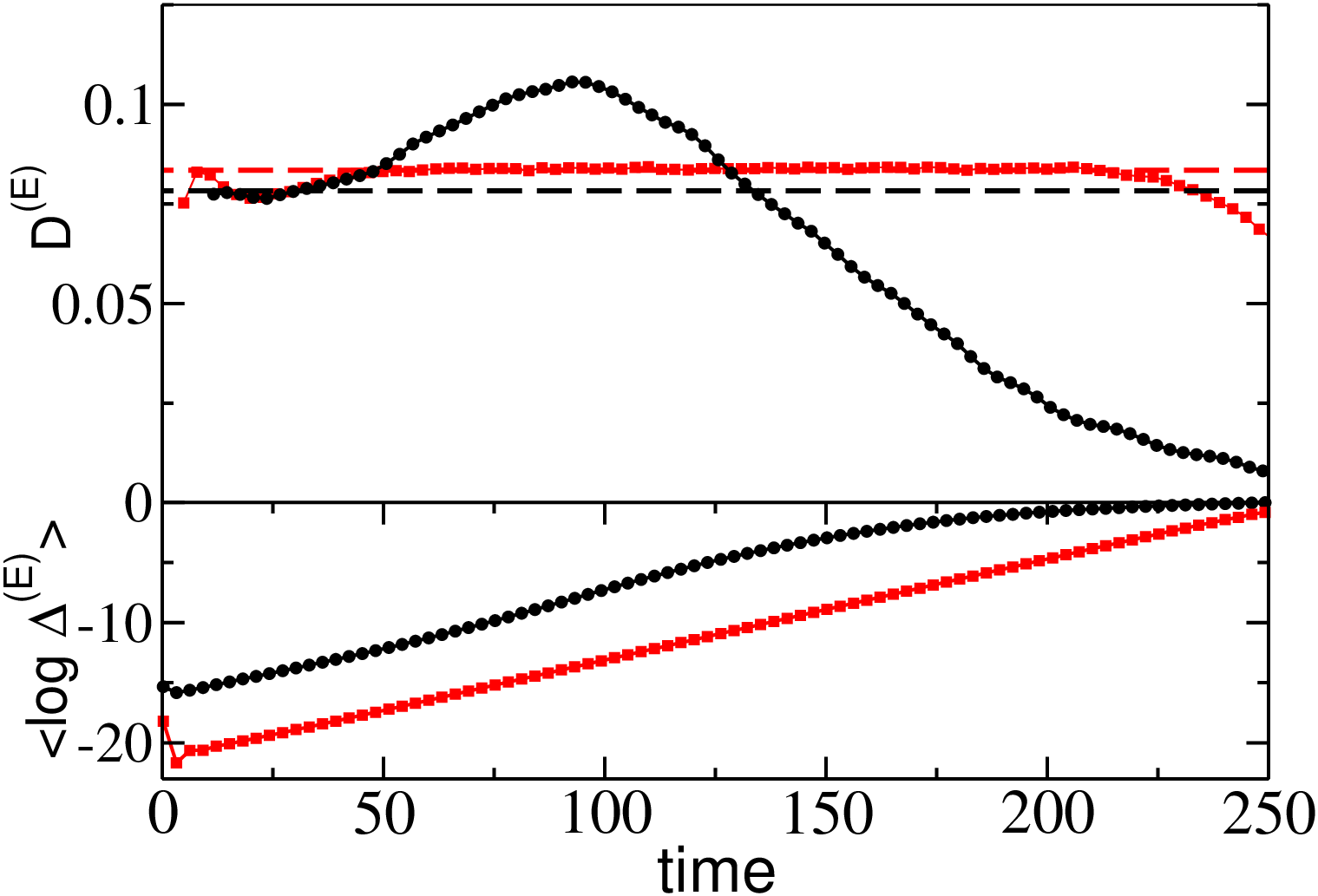}
\caption{Lower panel: Evolution of the average distance $<\log \Delta^{(E)}>$ as a function of time, for the 
current (red square) and the fluctuation (black circle) driven cases. The curves  
are obtained by averaging the distances between the perturbed and unperturbed trajectories over 5000 
(15000) realizations, after applying an initial perturbation of ${\cal O}(10^{-8})$. 
Upper panel: Indicator $D^{(E)}$ as a function of time for the same cases, calculated as the time derivative 
of $<\log \Delta^{(E)}>$. For small perturbations, $D^{(E)}$ is close to $\lambda$ (thick dashed lines), while 
observing a finite size effect is observable in the fluctuation driven case. 
The current and fluctuation driven cases have been examined
for the same parameter values reported in Fig. \ref{fig:FSLE}.
}
\label{fig:lyapunovVStime}
\end{figure}

\begin{figure}[htb]
\centering
\includegraphics[width=0.6\textwidth ]{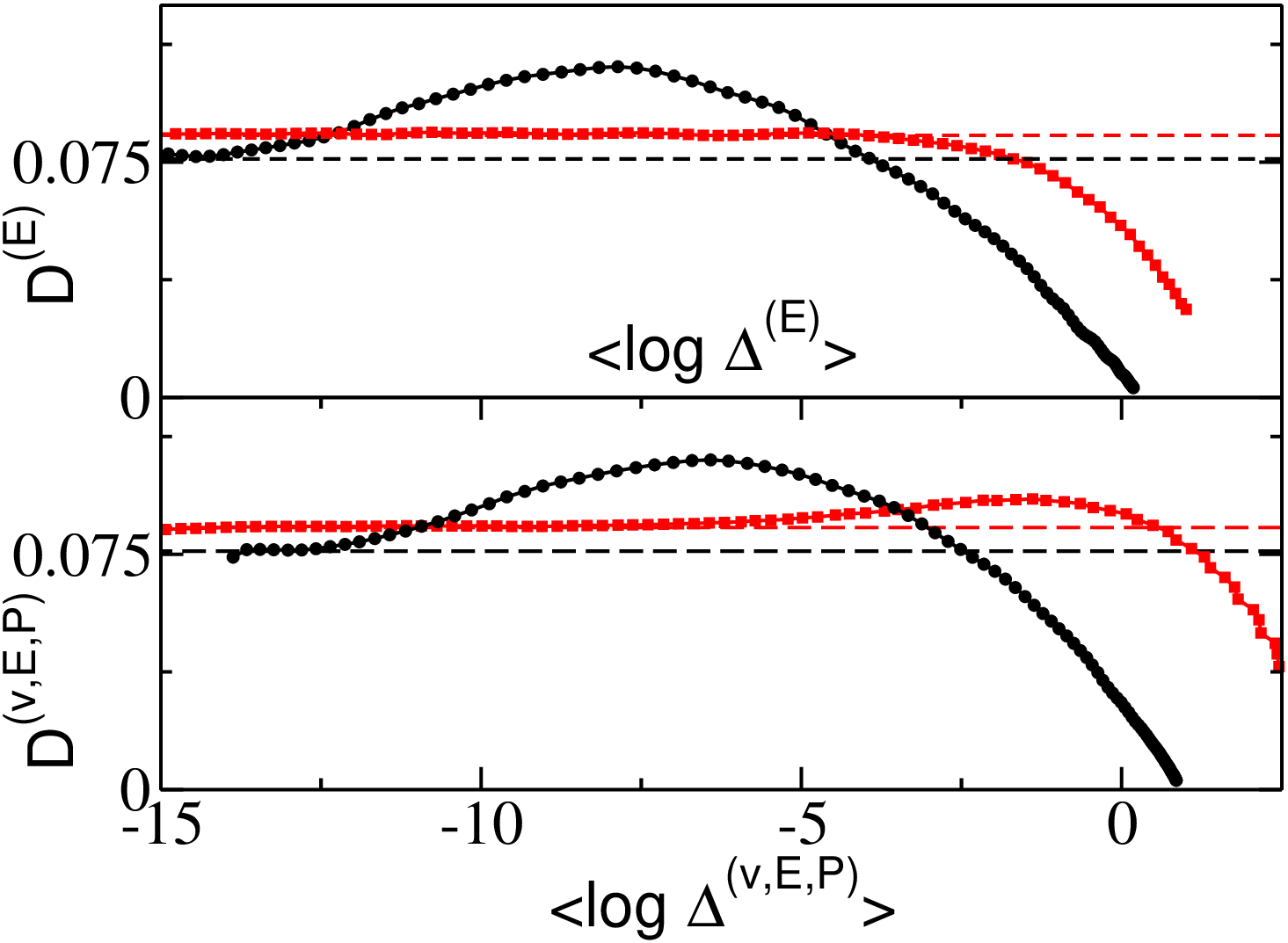}
\caption{Indicator $D(\Delta)$ versus the complete distance $\Delta^{(v,E,P)}$ (lower panel)
and versus the distance $\Delta^{(E)}$ (upper panel)
for the current (red squares) and fluctuation (black circles) driven cases.
 The curves are obtained with the same procedure described in the caption of Fig. \ref{fig:lyapunovVStime}. In both 
panels, thick dashed lines illustrate the corresponding value of $\lambda$. 
The current and fluctuation driven cases have been examined
for the same parameter values reported in Fig. \ref{fig:FSLE}.
}
\label{fig:lyapunovVSlog_TOTAL}
\end{figure}

The estimation of the FSLE, as
already mentioned, suffers of several numerical problems in these systems.
Therefore we decided to consider the indicator $D(\Delta^{(E)}(t))$, for simplicity denoted as $D^{(E)}$, which is
less affected by the single orbit fluctuations, since its estimation is based on the
time derivative of the averaged distance $\left\langle \log  \Delta (t)  \right\rangle$.
In Fig.~\ref{fig:lyapunovVStime} we report $\left\langle \log  \Delta^{(E)} (t)  \right\rangle$
and $D^{(E)}$ as a function of time for a current driven and a 
fluctuation driven case, in both situations after an initial transient, 
the indicator $D^{(E)}$ coincides with $\lambda$.
However, in the current driven case it coincides with $\lambda$ for a very long time
before decreasing due to the folding of the trajectories, while in the fluctuation
driven situation it becomes soon larger than the maximal Lyapunov exponent and it shows a 
clear peak at finite amplitudes, before the folding effect sets in.
 The same results are reported in the upper panel of  Fig.~\ref{fig:lyapunovVSlog_TOTAL}
as a function of $\left\langle \log  \Delta^{(E)} (t)  \right\rangle$, the peak
in the fluctuations driven case is located around $4 \times 10^{-4}$ thus
at a smaller amplitude with respect to the FSLE, despite the system size and parameters
are the same in both cases. Furthermore, in the lower panel in Fig.~\ref{fig:lyapunovVSlog_TOTAL} we report
the indicator $D(\Delta^{v,E,P}(t))$ ($D^{(v,E,P)}$ from now on) estimated for the total distance 
among the perturbed and unperturbed orbit. As expected, the discontinuities
present in the evolution of the membrane potentials and of the auxiliary field
$P$ due to pulse emission and pulse arrival, induce a small increase on $D^{(v,E,P)}$
with respect to the infinitesimal value $\lambda$ at finite amplitudes
even in the current driven case. However, in this case the peak of $D^{(v,E,P)}$
is definitely smaller with respect to the one observed in the fluctuation
driven case and it is located at larger perturbations ${\cal O}(1)$. 
Similar effects are observable also by considering the FSLE associated to 
$\Delta^{(v,E,P)}$, data not shown.  Nevertheless, in
order to keep ourselves in a consistent framework, in what follows we will consider 
the distance between the perturbed and unperturbed continuous fields $\Delta^{(E)}$. By choosing
this norm, we will avoid the presence of (trivial) peaks due to discontinuities as in the current driven
system, but instead, the presence of these peaks will be a genuine indication of 
nonlinear instabilities, as those present in a fluctuation driven regime.

\begin{figure}[htb]
\centering
\subfigure{\label{fig:DDD9}\includegraphics[width=0.45\textwidth ]{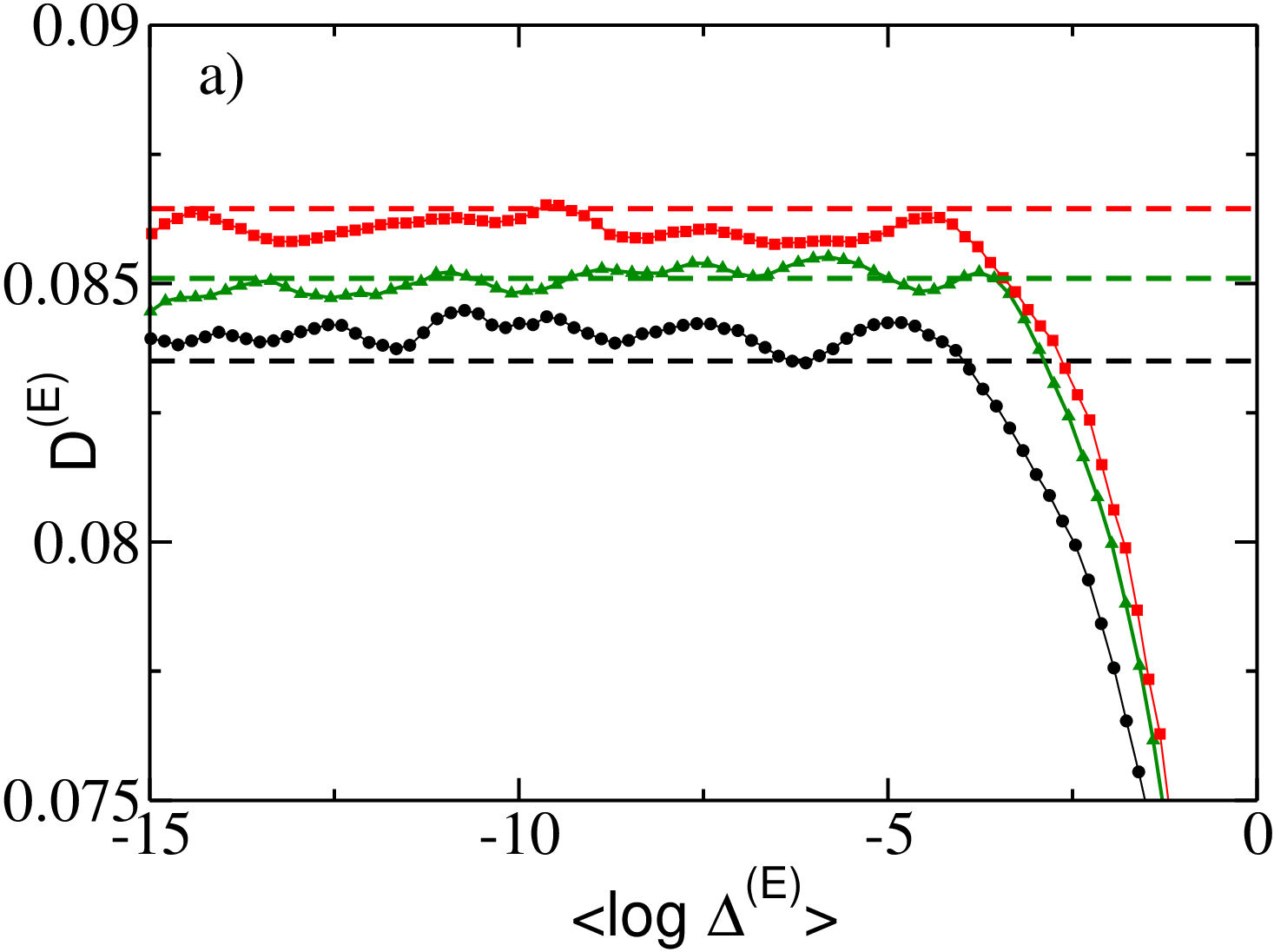}}
\subfigure{\label{fig:DDD3}\includegraphics[width=0.45\textwidth ]{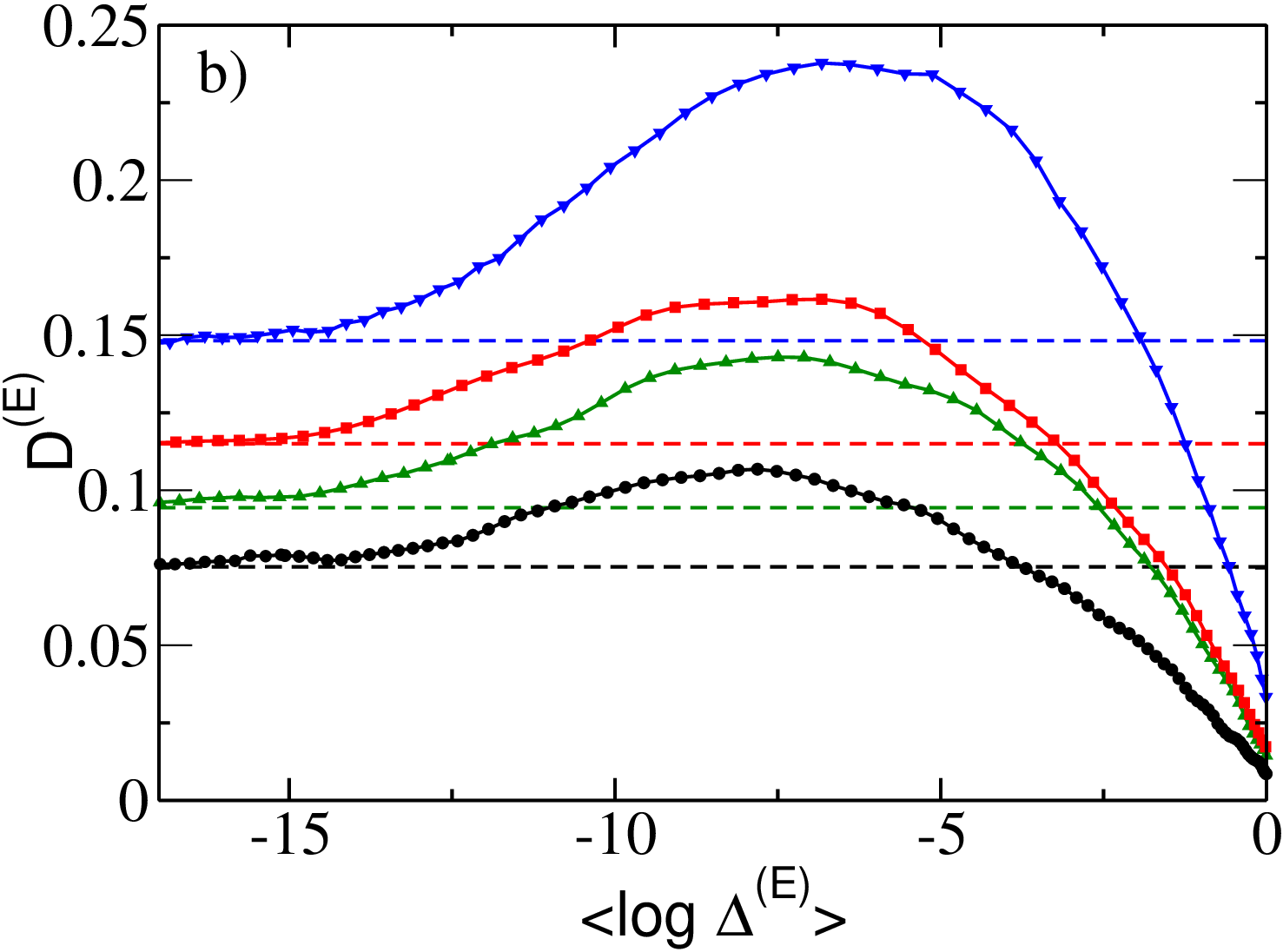}}
\caption{Finite amplitude perturbation analysis for several sizes of the network by using the procedure described in Fig. \ref{fig:lyapunovVSlog_TOTAL} for the distance $\Delta^{E}$ for the current (a) and fluctuation (b) driven
setups. In both cases the studied sizes correspond to $N=400$ (black circles), $N=600$ (green up-triangles), 
$N=800$ (red squares) and $N = 1600$ (blue down-triangles), averaged through 7500 realizations. 
Remaining parameters as in Fig. \ref{fig:FSLE}.
}
\label{fig:DDD}
\end{figure}

The indicator  $D^{(E)}$ is reported in Fig.~\ref{fig:DDD} for various system
sizes, ranging from $N=400$ to $N=1600$ for the current and fluctuation driven cases.
We observe that in the current driven case $D^{(E)}$ always gives a value around  
the corresponding $\lambda$ at all scales, apart the final saturation 
effect (see Fig.~\ref{fig:DDD9}). Notice that $\lambda$, for these system sizes, 
strongly depends on $N$ (as shown in \cite{Luccioli2012PRL}), the saturation at
the asymptotic value is expected to occur for $N > 5000$. For the fluctuation driven set-up,
a peak (larger than $\lambda$) is always present in $D^{(E)}$ at finite amplitudes 
(see Fig.~\ref{fig:DDD3}). The peak broadens for increasing $N$ extending to larger
amplitudes and also its height increases. The presence of more neurons in the
network renders stronger the finite amplitude effects, while nonlinear 
instabilities are present at larger and larger perturbation amplitudes.

 So far we have considered only chaotic regimes, both in the fluctuation
driven and in the current driven case. However, 
even in linearly stable cases the dynamics can be
erratic, as shown in Fig.~\ref{fig:rasters} for the fluctuation driven case corresponding
to $\alpha=5$ for which the maximal Lyapunov is negative at any system size (see Fig.~\ref{fig:lyapunov} (b)).
This kind of erratic behavior, known as Stable Chaos \cite{politi2010stable}, is one the most striking examples of dynamics 
driven by nonlinear effects, since the linear instabilities are asymptotically damped. 
In this situation neither the FSLE nor the indicator $D(\Delta)$ can be measured.
The reason is that, in order to ensure that the dynamics will take place on the associated attractor,
finite amplitude perturbations are reached only by starting from 
very small initial perturbations, which in this case are damped. Therefore, we should employ different 
indicators, namely the finite amplitude growth rate $R_T (\Delta_0)$ and the probability $P_S(\Delta_0)$.
 
\begin{figure}[htb]
\centering
\subfigure{\label{fig:LK_alpha5}\includegraphics[width=0.45\textwidth ]{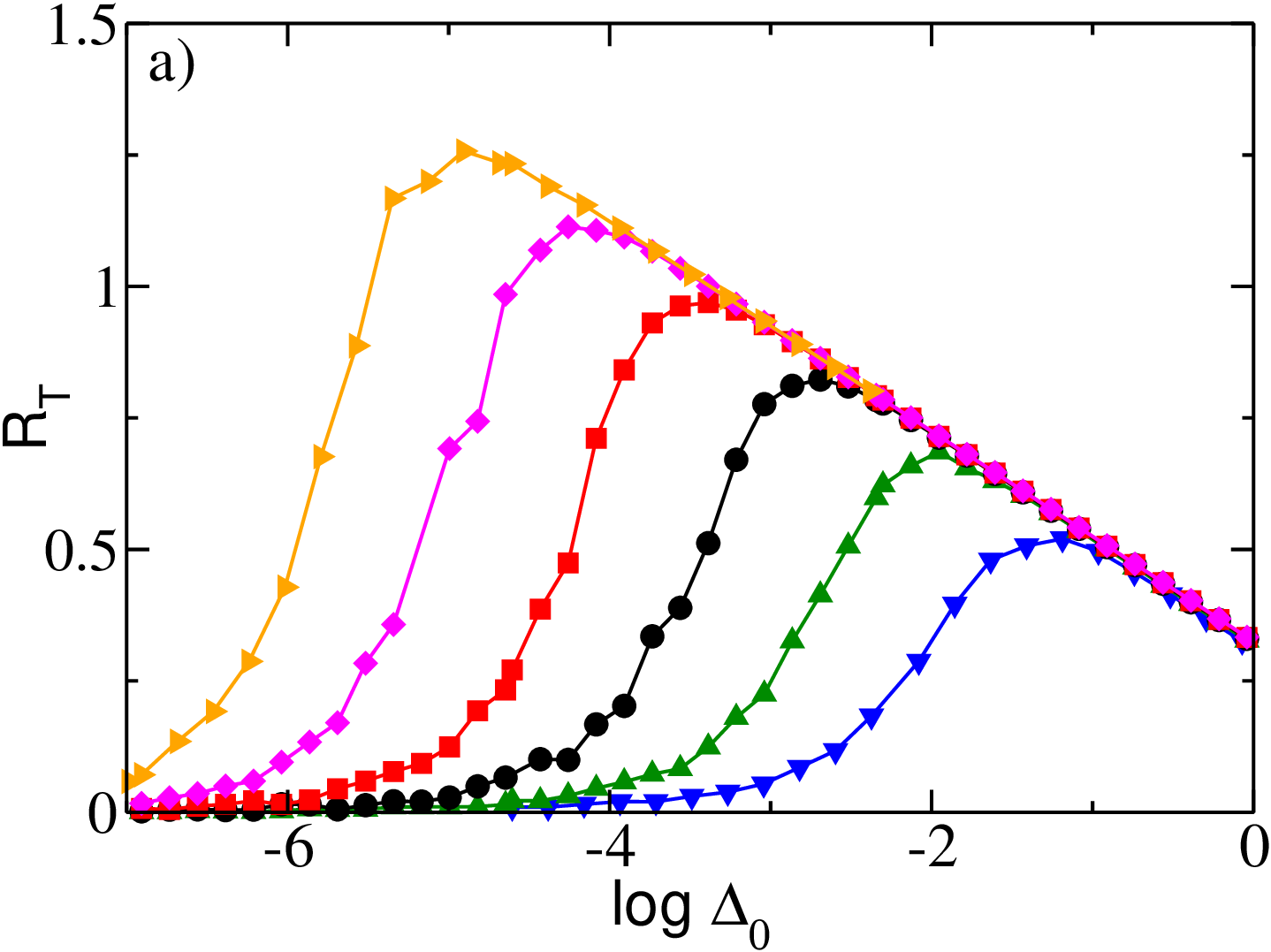}}
\subfigure{\label{fig:ProbTrans}\includegraphics[width=0.45\textwidth ]{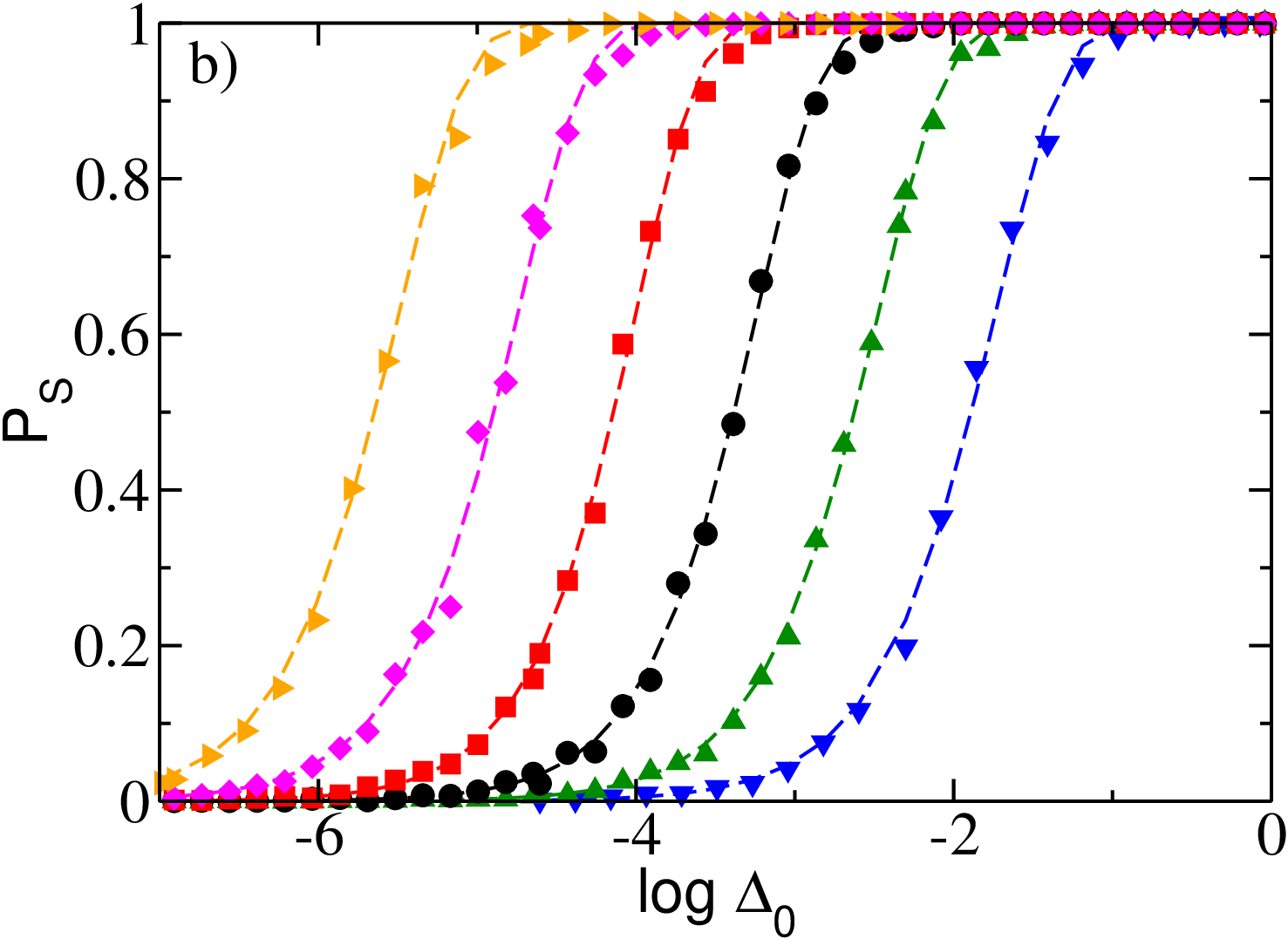}}
\caption{Characterization of the Stable Chaos regime: finite amplitude instabilities for different network sizes.
(a) $R_T$ indicator as a function of the initial perturbation
$\Delta_0$. (b) Probability $P_S$ to observe an exponential increase of the distance 
between a perturbed and an unperturbed orbit versus the initial perturbation
$\Delta_0$. Thick dashed lines refer to the fit to the data with the expression
$P_S = 1 - e^{-(\Delta_0/\beta)^{\mu}}$.
The studied sizes are $N=100$ (blue
down-triangle), $N=200$ (green up-triangles), $N=400$ (black circles), $N = 800$
(red squares), $N=1600$ (magenta diamonds) and $N=3200$ (orange right-triangles). For each 
perturbation $\Delta_0$, $R_T$ and $P_S$ are
calculated after $T = 5$ time units, threshold defining expanding trajectories $\theta_L = -2$ and averaging over $N_T = 5000$ realizations. Remaining parameters as reported in Fig.~\ref{fig:rasters}.}
\label{fig:RT_descrip}
\end{figure}

\begin{figure}
\centering
\subfigure{\label{fig:PeaksVsN}\includegraphics[width=0.45\textwidth ]{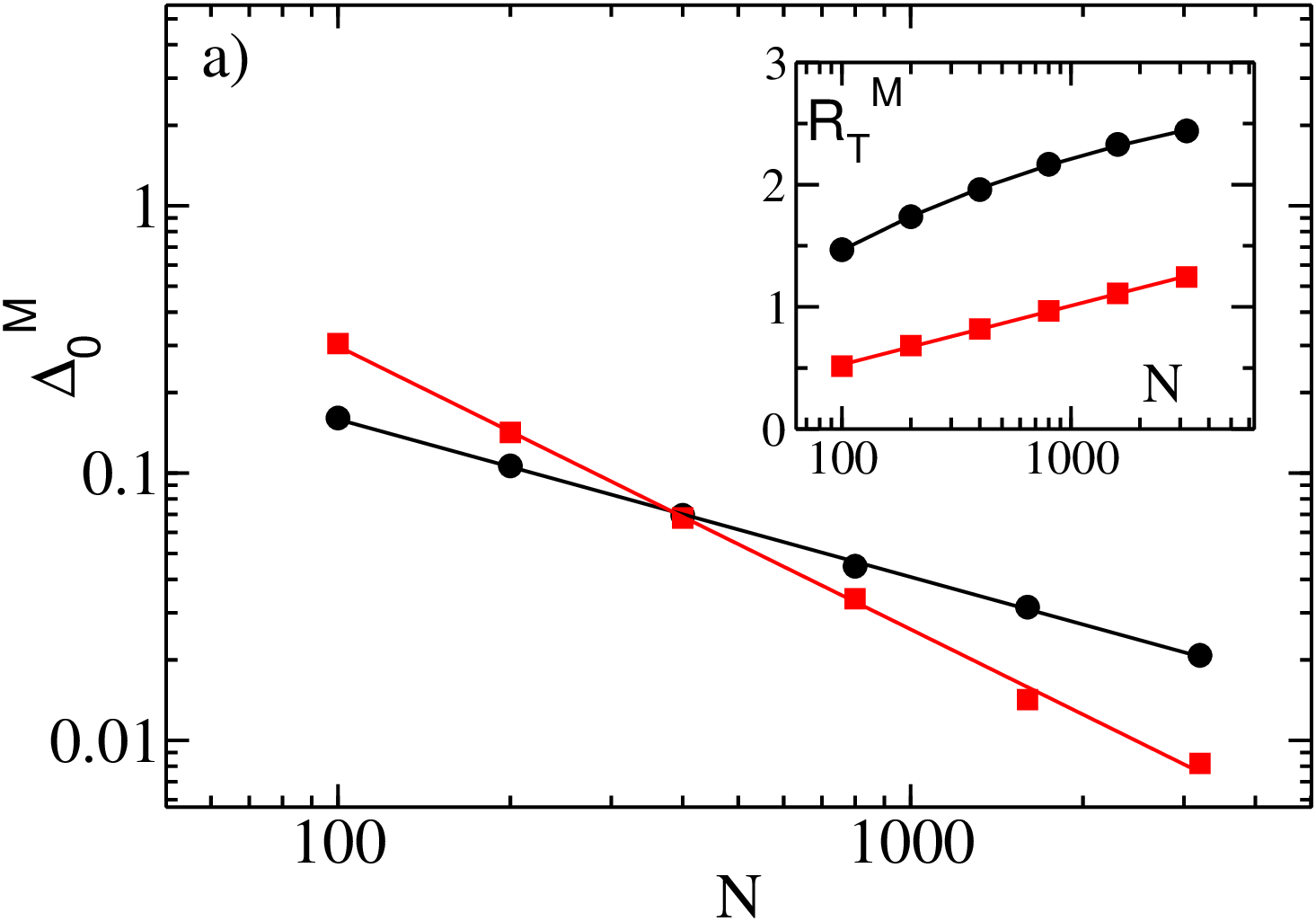}}
\subfigure{\label{fig:betaVsN}\includegraphics[width=0.45\textwidth ]{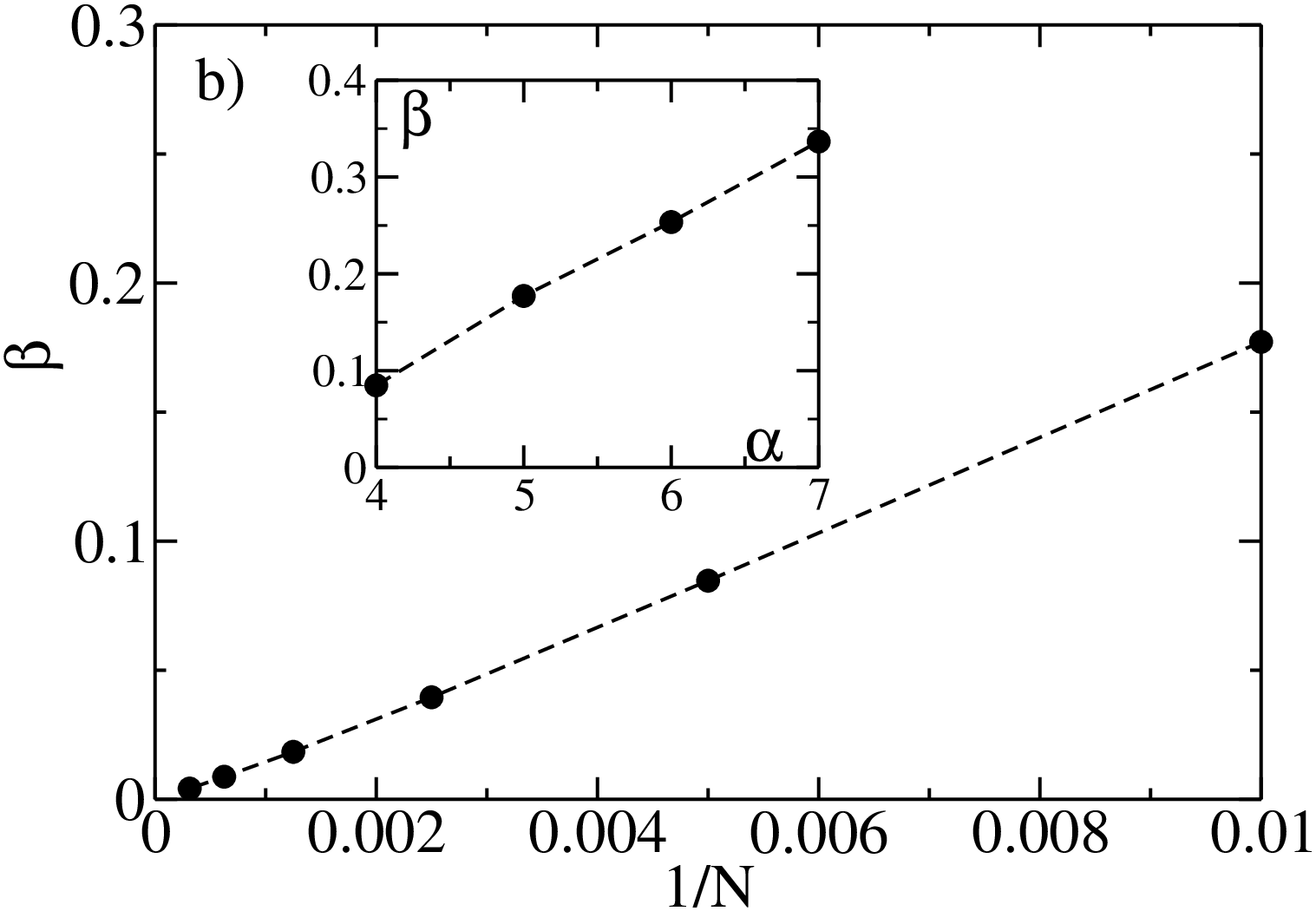}}
\caption{a) Peak position $\Delta_0^M$ as a function of $N$ in a log-log
scale for $\alpha = 3$ (black circles) and $\alpha = 5$ (red squares). The continuous
line are power law fitting $\Delta_0^M \propto N^{-\Phi}$ , with exponents
$\Phi = -0.59$ ($\Phi = -1.05$) for $\alpha = 3$ ($\alpha = 5$).
Inset, maximum value of $R_T$ as a function of the number of neurons in the network $N$
in a log-lin scale. The solid lines refer to fittings to the data, namely
$R^M_T = 3.09 - 5.60 N^{-0.27}$ for $\alpha = 3$;
$R_T^M = 0.23 + 0.28 \log (N)$ for $\alpha = 5$. $R_T$ calculated after a time span $T = 1$ $(T = 5)$
for $\alpha = 3$ $(\alpha = 5)$.
b) Amplitude scale $\beta$ associated to the indicator $P_S$ as a function of $1/N$.
In the inset, $\beta$ is reported as a function of $\alpha$ for parameter
values associated to non chaotic dynamics. In the same range
the exponent $\mu \sim 2.32$ (not shown). The model parameters refer to the 
fluctuation driven case studied in Fig.~\ref{fig:rasters}. Inset is obtained with $N = 100$}
\label{fig:TransitionParams}
\end{figure}

 As shown in Fig.~\ref{fig:LK_alpha5}, for the linearly stable fluctuation driven case
corresponding to $\alpha=5$, $R_T (\Delta_0) \to 0 $ for sufficiently small perturbations, as expected.
However $R_T (\Delta_0)$ becomes soon positive for finite amplitude perturbation and it reveals a large
peak $R_T^M$ located at an amplitude $\Delta_0^M$. For increasing system size $N$, as shown in 
Fig. \ref{fig:PeaksVsN} a linear decrease
of $\Delta_0^M$ with $N$ is clearly observable, while  $R_T^M$ reveals a logarithmic
increase with $N$. Thus suggesting that this indicator will diverge to
infinity in the thermodynamic limit, similarly to the results previously reported 
in~\cite{vanVreeswijk1996,MonteforteBalanced2012}. However, 
at variance with these latter studies, in the present context the connectivity remains finite even in the limit $N \to \infty$.

The analysis of $P_S(\Delta_0)$, reported in Fig.~\ref{fig:ProbTrans}, reveals that 
the curve can be well fitted as
\begin{equation} 
P_S(\Delta_0) = 1 - \exp(-\Delta_0/\beta)^\mu \qquad;
\label{fitPS}
\end{equation}
analogously to what done in \cite{MonteforteBalanced2012}. The parameter $\beta$ can be considered
as a critical amplitude, setting the scale over which nonlinear instabilities take place.
At variance with the results reported by Monteforte \& Wolf in~\cite{MonteforteBalanced2012}, 
we observe a linear decrease with $N$ of the critical amplitude $\beta$ (see Fig. \ref{fig:betaVsN})
and an exponent $\mu \simeq 2.3- 2.5$, depending on the employed system size.
Instead, Monteforte \& Wolf reported a scaling $\beta \propto 1/\sqrt{N}$
and an exponent $\mu = 1$. Furthermore, we have verified for various continuous $\alpha$ pulses, with
$\alpha \in [4;7]$, that the measured exponent $\mu$ does not particularly depend
on $\alpha$. The model here studied differs for the shape of the post-synaptic currents
from the one examined in~\cite{MonteforteBalanced2012}, 
where $\delta$-pulses have been considered.

   In our opinion, these two latter indicators, $R_T$ and $P_S$ bear essentially the same information:
they measure the propensity of a perturbation $\Delta_0$ to be amplified on a short time scale $T$.
This is confirmed by the fact that (as shown in Fig. \ref{fig:TransitionParams}) the values of $\Delta_0^M$ and
$\beta$, which set the relevant amplitude scales for the two indicators, both decrease with the same scaling
law (namely, $1/N$) with the system size. A possible explanation for this scaling could be  found
by assuming that the main source of nonlinear amplification is associated to a spike removal (addition)
in the perturbed orbit. A missing (extra) spike will perturb, to the leading order,
the distance $\Delta^{(v,EP)}$ by an amount $\propto \alpha^2\sqrt{K}/N$, since the lost (added) 
post-synaptic pulses are $K$ each of amplitude $\alpha^2/\sqrt{K}$.
This argument explains as well the logarithmic increase of $R_T^M$ with the system size.
Furthermore, the decrease of $\Delta_0^M$ and $\beta$ with $N$ seems to indicate that 
in the thermodynamic limit any perturbation, even infinitesimal, 
will be amplified. 
This is clearly in contradiction with the fact that  the system is linearly stable and it appears to remain stable by increasing $N$
(as shown in Fig. \ref{fig:lambda_vs_N}). In systems exhibiting Stable Chaos, it has been reported many times
the fact that the thermodynamic limit and the infinite time limit do not commute~\cite{politi1993}.
For finite system size, at sufficiently large times (diverging exponentially with $N$)
a stable state is always recovered, while if the thermodynamic limit is taken before
the infinite time one, the system will remain erratic at any time~\cite{politi2010stable}.
In the present case, it seems that a different non commutativity between
the thermodynamic limit and the limit of vanishingly small perturbations is present,
similar conclusions have been inferred also in~\cite{MonteforteBalanced2012}.
Therefore, we can apparently conclude that a fluctuation driven system,  which is linearly 
stable, but presents nonlinear instabilities, will become unstable at any 
amplitude and time scales in the thermodynamic limit.
However, one should be extremely careful in deriving any conclusion from these
indicators, since they are not dynamical invariant and their values depend not only
on the considered variables but also on the employed norm. 
Furthermore, in the present context there is an additional problem related to
the meaningful definition of the norm in an infinite space, as that achieved in the 
thermodynamic limit.

To understand the limit of applicability of $R_T$, we have examined this indicator
also in the chaotic fluctuation driven case, namely for $\alpha=3$.
Also in this case we observe that $\Delta_0^M$ will vanish for diverging
system size, but with a different scaling law, namely 
$\Delta_0^M \simeq N^{-0.6}$. Furthermore,  $R_T^M$ increases with $N$,
but this time it appears to saturate in the thermodynamic limit 
with a scaling law similar to the one reported in (\ref{eq:fitting}) for the
maximal Lyapunov exponent, more details are reported in the caption of Fig. \ref{fig:TransitionParams}.
Unlike the stable regime, in the chaotic one we cannot justify 
with the simple spike addition (removal) argument
the scaling with $N$ neither for $\Delta_0^M$ nor for $R_T^M$. 
It is high probable that in this regime the interactions of the linear and 
nonlinear instabilities leads to more complicated mechanisms. 
The evolution of the indicator $R_T$ suggests that for increasing $N$ its 
peak will move down to smaller and smaller amplitude scale. However, this result is
in contradiction with the behavior of $D^{(E)}$ reported in Fig.~\ref{fig:DDD3},
for this latter indicator the position of the peak is not particularly affected by $N$.
In particular, finite amplitude instabilities affect larger and larger scales,
contrary to what seen for $R_T$ (see Fig.~\ref{fig:LK_alpha5}). 
The same behavior is observable for $D^{(v,E,P)}$, data not shown.
These contradictory results point out the limits of indicators like $R_T$ and
$P_S$ relying on dynamical evolutions not taking place on the attractor of the system.

\begin{figure}
\centering
\subfigure{\includegraphics[width=0.45\textwidth ]{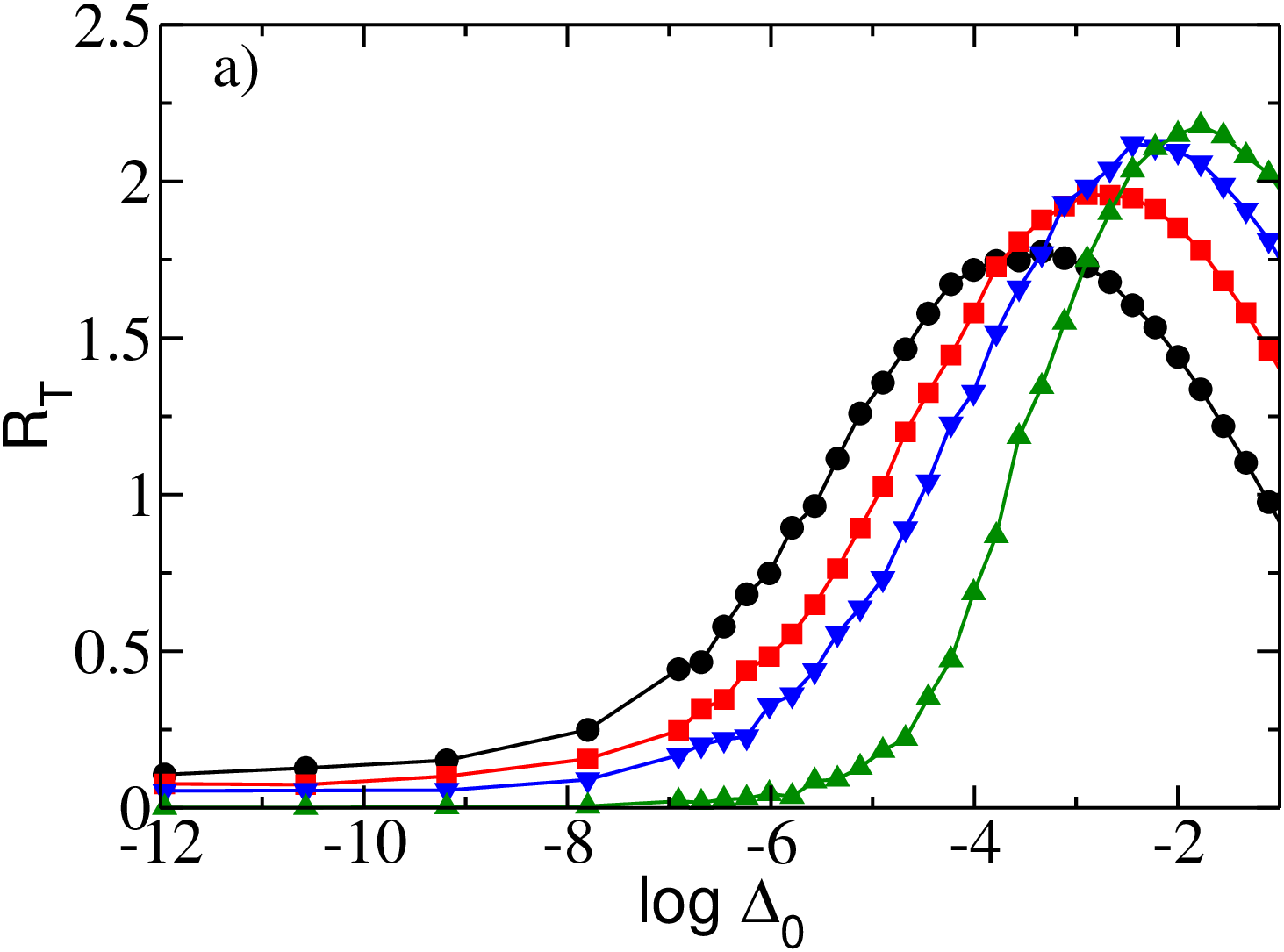}}
\subfigure{\includegraphics[width=0.45\textwidth ]{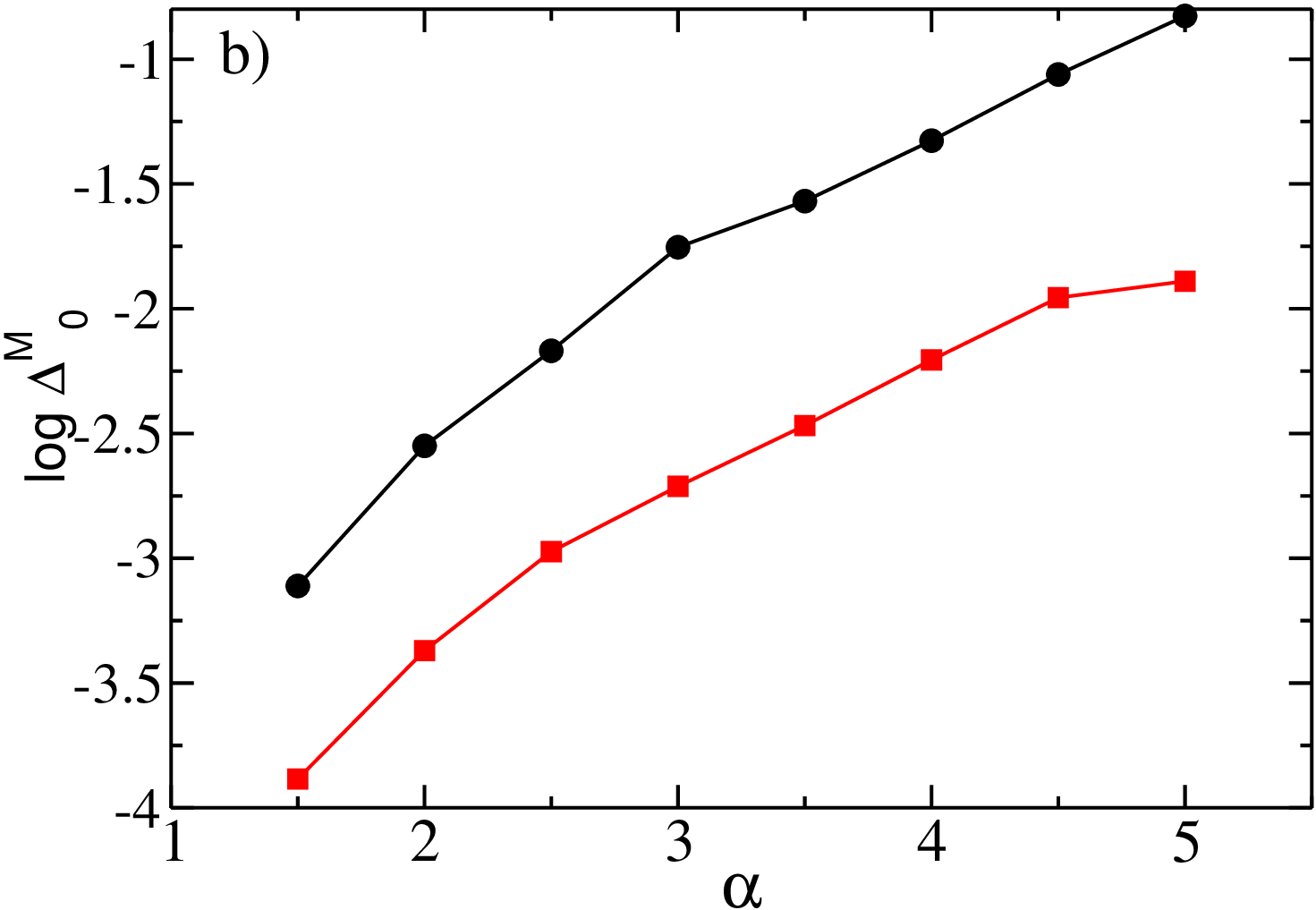}}
\caption{Finite size instabilities for fluctuation driven dynamics,
for different pulse widths. 
(a) $R_T$ as a function of the initial perturbation $\Delta_0$, for $\alpha = 2$ (black circle), 
$\alpha = 3$ (red square), $\alpha = 4$ (blue down-triangle), $\alpha = 5$ (green up-triangle). 
The system size is fixed to $N=400$.
 (b) Peak location in logarithmic scale $\log \Delta_0^M$ as a function of 
the inverse pulse-width $\alpha$, for two sizes of the network: $N=100$ (black circles) and $N=400$ (red squares).
The peak positions were found by fitting a quadratic function around the maximum of the
function $R_T$ in (a). $R_T$ is calculated as described in the caption of Fig.~\ref{fig:RT_descrip}. 
Remaining parameters as in Fig. \ref{fig:rasters}.
\label{fig:PeakLocation}}
\end{figure}

Finally, in order to study the effect of the  pulse shape on the finite amplitude behavior as measured by
$R_T$, we proceeded to calculate this indicator for various $\alpha$-values. As shown 
in Fig.~\ref{fig:PeakLocation}, 
for increasing $\alpha$ (corresponding to narrower peaks) the position of the maximum $\Delta_0^M$ moves towards 
larger amplitudes. This effect can be explained by the fact that the maximal Lyapunov exponent decreases 
with $\alpha$ (as shown in Fig.~\ref{fig:lambda_vs_alpha})
and therefore perturbations of bigger and bigger amplitudes are required to 
destabilize the system for vanishingly pulse width. Consistently also the
parameter $\beta$ associated to $P_S$ increases for increasing $\alpha$-values,
as shown in the inset of Fig.~\ref{fig:TransitionParams} (b).

\section{Discussion}
\label{sec:discussion}

We have investigated the dynamics and stability of current and fluctuation driven
neural networks, the former (latter) have been realized as a purely excitatory (inhibitory)
pulse coupled network of leaky integrate-and-fire (LIF) neurons with a sparse architecture. In particular, 
we considered random networks with a constant in-degree $K=20$ for any examined size.

The excitatory network, despite being chaotic, reveals a low spiking variability. On the other hand, 
in the fluctuation driven case the variability is high for any considered pulse width and system size
(CV $\simeq 0.7 -1.0$). However, a different picture arises when studying the stability 
of infinitesimal perturbations: the system is chaotic for slow synapses and 
it becomes stable for sufficiently fast synaptic times ($\le 4$ ms).
Furthermore, a chaotic state for the inhibitory network
is observable already at small connectivity $K \sim O(10^1)$ contradicting
what reported in~\cite{Zillmer2009LongTrans}, where the authors affirmed that
a large connectivity is a prerequisite to observe chaotic motion in these models.

The maximal Lyapunov exponent $\lambda$ tends towards an asymptotic value for increasing
system sizes with a power-law scaling. The exponent $\eta$ associated to this scaling
is different in the current (fluctuation) driven case, in particular $\eta \simeq 1$ 
($\eta \simeq 1/4$)~\cite{Luccioli2012PRL}. In the fluctuation driven situation the exponent is the
same in the chaotic and stable phases. The origin of the observed scaling demands
for new theoretical analysis, similar to the one developed for spatio-temporal chaotic
systems~\cite{pikovsky1998dynamic}.

Quite astonishingly even in the linearly stable regime an erratic evolution of the network
is observable. A similar phenomenon has been already observed in several systems
ranging from diffusively coupled chaotic maps to neural networks, and it has
been identified as stable chaos~\cite{politi2010stable}. In this context,
finite amplitude perturbations are responsible for the erratic behavior observed in the
system. In diffusively coupled systems this nonlinear instabilities
has been characterized in terms of the propagation velocity of the information
and of suitable Finite Size Lyapunov Exponents (FSLEs)~\cite{Torcini1995FSLE, CenciniTorcini}.
FSLEs have been previously employed in the context
of fully coupled neural networks, where they revealed that the origin
of the chaotic motion observed in two symmetrical coupled neural populations was
due to collective chaos in the mean-field variables driving the single LIF
neurons~\cite{Olmi2010Oscillations}. In the context of randomly coupled systems the 
concept of propagation velocity on a lattice looses his sense, while FSLEs reveal serious 
problems in their numerical implementation. 

However, FSLEs clearly show also in our case that in the current driven case
the observed instabilities have a purely linear origin, while
in the fluctuation driven situation nonlinear mechanisms are present
even when the system is chaotic. This analysis is confirmed by
a novel indicator we have introduced, namely the local derivative $D(\Delta)$ 
of the averaged logarithmic distance $< \log \Delta >$  between the reference and the perturbed
trajectory. This quantity suffers less than the FSLE the trial to trial fluctuations,
since it is based on an averaged profile. For the fluctuation driven case this indicator
is larger than the maximal Lyapunov exponent at finite amplitudes and this
effect is present for all the examined system sizes. The position of the peak in $D(\Delta)$ 
seems not to be particularly influenced by the system size, while the peak itself
broadens towards larger amplitudes for increasing $N$.
Unfortunately, all these indications cannot tell us if the nonlinear mechanisms
are prevailing on the linear ones, but just that the nonlinear effects are present.
To measure the influence of linear versus nonlinear effects on the system dynamics,
novel indicators are required, similar to linear and nonlinear information velocities
for diffusively coupled systems~\cite{politi2010stable}.

 As a final point we have studied nonlinear instabilities in linearly stable systems
emerging in fluctuation driven inhibitory networks for sufficiently narrow postsynaptic
currents. For the characterization of these instabilities we have employed the average
finite amplitude growth rate $R_T(\Delta_0)$, measured after a finite time interval $T$,
analogously to what what done in ~\cite{LetzKantz2000, CenciniTorcini}, 
and the probability $P_S(\Delta_0)$ that an initial perturbation induces an
exponential separation between the perturbed and the reference orbits, previoulsy
introduced in~\cite{MonteforteBalanced2012}. Both these indicators reveal the
existence of instabilities associated to finite perturbations, in particular 
the characteristic amplitude scales associated to these indicators vanish
in the thermodynamic limit as $1/N$. Thus suggesting that instabilities in these
systems can occur even for infinitesimal perturbations in clear contradiction
with the fact that these systems are linearly stable at any system size, as revealed 
by the Lyapunov analysis. This contradiction has lead Monteforte \& Wolf to conjecture in~\cite{MonteforteBalanced2012}
that the thermodynamic limit and the limit of of vanishingly small perturbations
do not commute in these models. Furtermore, we measure a logarithmic divergence with the system size
of the peak height of $R_T(\Delta_0)$, suggesting that in the thermodynamic limit
the value of these indicator will become infinite, similarly to what found 
in the high connectivity limit for a
binary neuronal model in the balanced state~\cite{vanVreeswijk1996}
and for LIF with $\delta$-pulses in ~\cite{MonteforteBalanced2012}. However,
in our study the connectivity remains finite and small in the limit $N \to \infty$.

Our opinion, based on the comparison
of the indicators $D(\Delta)$ and $R_T(\Delta_0)$ performed in a
fluctuation driven chaotic situation, is that the above results can be
due to the fact that the dynamics considered for the estimation of 
$R_T(\Delta_0)$ and $P_S(\Delta_0)$ do not take place on the attractor 
of the system. This because the indicators are estimated at short times,
without allowing the perturbed dynamics to relax onto the attractor.
The development of new indicators is required to analyze more in depth
the phenomenon of Stable Chaos in randomly connected networks.

\section*{Acknowledgments}
We thank A. Politi, S. Luccioli, and J. Berke for useful  discussions. 
This work has been supported by the European Commission 
under the program ``Marie Curie Network for Initial Training", through the
project N. 289146, ``Neural Engineering Transformative Technologies (NETT)". D. A.-G. would like also to acknowledge
the partial support provided by ``Departamento Adminsitrativo de Ciencia Tecnologia e Innovacion - Colciencias" through
the program ``Doctorados en el exterior - 2013".
\\

\section*{Appendix I}

The LIF model is usually expressed in physical units as follows~\cite{Koch1998} 
\begin{equation}
\tau_m \frac{dV}{d\hat{t}} = -(V(\hat{t}) - V_0) + R_m I_{ext} + \tau_m
\hat{g} \hat{E} (\hat{t})  \qquad ,
\label{physical}
\end{equation}
where $R_m$ is the specific mebrane resistance, $\tau_m$
the membrane time constant, and $V_0$ the resting potential.
The transformation of the adimensional model (\ref{eq:mebrane}) to 
(\ref{physical}) can be obtained by performing the following set of transformations
\begin{eqnarray}
V_i=&v_i (V_{th}-V_0)+V_0 \qquad R_m I_{ext}=&a (V_{th}-V_0)+V_0 \\
\hat{g}=&g(V_{th}-V_0)+V_0 \qquad \hat{t} =& t\tau_m \qquad ,
\end{eqnarray}
where $V_{th}$ is the firing threshold value. Notice that $\hat{\alpha} = \alpha/\tau_m$ and
$\hat{E} = E /\tau_m$ have the dimensionality of a frequency and $\hat{g}$
of a potential. Realistic values for the introduced parameters, are $\tau_m = 20$ ms,
$V_0 = -60$ mV, $V_{th} = -50$ mV \cite{Sterratt2011Principles}.

The postsynaptic current rise times $1/\hat{\alpha}$ employed in this article
range from $4$ to $20$ ms for inhibitory cells, while it is fixed to 2.22 ms for excitatory ones.
Furthermore, the  average neuronal firing rates are of order $\simeq 50$ Hz ($\simeq 6$ Hz)  
for excitatory (inhibitory) networks, which are quite reasonable values for pyramidal  neurons
(inter-neurons) of the cortex~
\cite{mccormick1985comparative,SoftkyKoch1993,destexhe2001fluctuating,Stiefel_Sejnowski2013OriginIrregular}






\section*{Bibliography}



\begin{thebibliography}{10}
\expandafter\ifx\csname url\endcsname\relax
  \def\url#1{\texttt{#1}}\fi
\expandafter\ifx\csname urlprefix\endcsname\relax\def\urlprefix{URL }\fi
\expandafter\ifx\csname href\endcsname\relax
  \def\href#1#2{#2} \def\path#1{#1}\fi

\bibitem{SoftkyKoch1993}
W.~R. Softky, C.~Koch, The highly irregular firing of cortical cells is
  inconsistent with temporal integration of random epsps, J. Neurosci. 13~(1)
  (1993) 334--350.

\bibitem{holt1996}
G.~R. Holt, W.~R. Softky, C.~Koch, R.~J. Douglas, Comparison of discharge
  variability in vitro and in vivo in cat visual cortex neurons, J.
  Neurophysiol. 75~(5) (1996) 1806--1814.

\bibitem{Stiefel_Sejnowski2013OriginIrregular}
K.~M. Stiefel, B.~Englitz, T.~J. Sejnowski, Origin of intrinsic irregular
  firing in cortical interneurons, P. Nat. Acad. Sci. USA 110~(19) (2013)
  7886--7891.

\bibitem{tuckwell2005}
H.~C. Tuckwell, Introduction to theoretical neurobiology: Volume 2, nonlinear
  and stochastic theories, Vol.~8, Cambridge University Press, 2005.

\bibitem{Abbott1993Asynchronous}
L.~F. Abbott, C.~van Vreeswijk, {Asynchronous states in networks of
  pulse-coupled oscillators}, Phys. Rev. E 48~(2) (1993) 1483--1490.

\bibitem{Haider2006BalancedExperiment}
B.~Haider, A.~Duque, A.~R. Hasenstaub, D.~A. McCormick, Neocortical network
  activity in vivo is generated through a dynamic balance of excitation and
  inhibition, J. Neurosci. 26~(17) (2006) 4535--4545.

\bibitem{renart2010}
A.~Renart, J.~de~la Rocha, P.~Bartho, L.~Hollender, N.~Parga, A.~Reyes, K.~D.
  Harris, The asynchronous state in cortical circuits, Science 327~(5965)
  (2010) 587--590.

\bibitem{litwin2012slow}
A.~Litwin-Kumar, B.~Doiron, Slow dynamics and high variability in balanced
  cortical networks with clustered connections, Nature. Neurosci. 15~(11)
  (2012) 1498--1505.

\bibitem{TsodyksSenojwski1995}
M.~V. Tsodyks, T.~Sejnowski, Rapid state switching in balanced cortical network
  models, Network-Comp. Neural. 6~(2) (1995) 111--124.

\bibitem{BrunelHakim1999}
N.~Brunel, V.~Hakim, Fast global oscillations in networks of integrate-and-fire
  neurons with low firing rates, Neural. Comput. 11~(1) (1999) 1621--1671.

\bibitem{NeuronalNetwroks2005Vogels}
T.~P. Vogels, K.~Rajan, L.~Abbott, Neural networks dynamics, Annu. Rev.
  Neurosci. 28~(1) (2005) 357--376.

\bibitem{Brunel2000Sparse}
N.~Brunel, Dynamics of sparsely connected networks of excitatory and inhibitory
  spiking neurons, J. Comput. Neurosci. 8~(3) (2000) 183--208.

\bibitem{politi2010stable}
A.~Politi, A.~Torcini, Stable chaos, in: Nonlinear Dynamics and Chaos: Advances
  and Perspectives, Springer, 2010, pp. 103--129.

\bibitem{Zillmer2006}
R.~Zillmer, R.~Livi, A.~Politi, A.~Torcini, Desynchronization in diluted neural
  networks, Phys. Rev. E 74~(3) (2006) 036203.

\bibitem{JhankeTimme2008PRL}
S.~Jahnke, R.-M. Memmesheimer, M.~Timme, Stable irregular dynamics in complex
  neural networks, Phys. Rev. Lett. 100 (2008) 048102.

\bibitem{Zillmer2009LongTrans}
R.~Zillmer, N.~Brunel, D.~Hansel, Very long transients, irregular firing, and
  chaotic dynamics in networks of randomly connected inhibitory
  integrate-and-fire neurons, Phys. Rev. E 79 (2009) 031909.

\bibitem{Timme2009ChaosBalance}
S.~Jahnke, R.-M. Memmesheimer, M.~Timme, How chaotic is the balanced state?,
  Front. Comp. Neurosci. 3~(13).

\bibitem{Luccioli2010Irregular}
S.~Luccioli, A.~Politi, {Irregular Collective Behavior of Heterogeneous Neural
  Networks}, Phys. Rev. Lett. 105~(15) (2010) 158104+.

\bibitem{MonteforteBalanced2012}
M.~Monteforte, F.~Wolf, Dynamic flux tubes form reservoirs of stability in
  neuronal circuits, Phys. Rev. X 2 (2012) 041007.

\bibitem{politi1993}
A.~Politi, R.~Livi, G.-L. Oppo, R.~Kapral, Unpredictable behaviour in stable
  systems, Europhys. Lett. 22~(8) (1993) 571.

\bibitem{tiesinga2000}
P.~Tiesinga, J.~V. Jos{\'e}, T.~J. Sejnowski, Comparison of current-driven and
  conductance-driven neocortical model neurons with hodgkin-huxley
  voltage-gated channels, Phys. Rev. E 62~(6) (2000) 8413.

\bibitem{renart2007}
A.~Renart, R.~Moreno-Bote, X.-J. Wang, N.~Parga, Mean-driven and
  fluctuation-driven persistent activity in recurrent networks, Neural. Comput.
  19~(1) (2007) 1--46.

\bibitem{London2010}
M.~London, A.~Roth, L.~Beeren, M.~H\"{a}usser, P.~E. Latham, Sensitivity to
  perturbations in vivo implies high noise and suggests rate coding in cortex,
  Nature. 466 (2010) 123--127.

\bibitem{Aurell1996FSLE}
E.~Aurell, G.~Boffetta, A.~Crisanti, G.~Paladin, A.~Vulpiani, Growth of
  noninfinitesimal perturbations in turbulence, Phys. Rev. Lett. 77 (1996)
  1262--1265.

\bibitem{turrigiano1998}
G.~G. Turrigiano, K.~R. Leslie, N.~S. Desai, L.~C. Rutherford, S.~B. Nelson,
  Activity-dependent scaling of quantal amplitude in neocortical neurons,
  Nature 391~(6670) (1998) 892--896.

\bibitem{turrigiano2008}
G.~G. Turrigiano, The self-tuning neuron: synaptic scaling of excitatory
  synapses, Cell 135~(3) (2008) 422--435.

\bibitem{Zillmer2007}
R.~Zillmer, R.~Livi, A.~Politi, A.~Torcini, Stability of the splay state in
  pulse-coupled networks, Phys. Rev. E 76 (2007) 046102.

\bibitem{Olmi2010Oscillations}
S.~Olmi, R.~Livi, A.~Politi, A.~Torcini, {Collective oscillations in disordered
  neural networks.}, Phys. Rev. E 81~(4 Pt 2).

\bibitem{Tattini2011Coherent}
L.~Tattini, S.~Olmi, A.~Torcini, Coherent periodic activity in excitatory
  erdös-renyi neural networks: the role of network connectivity, Chaos 22~(2)
  (2012) 023133.

\bibitem{Luccioli2012PRL}
S.~Luccioli, S.~Olmi, A.~Politi, A.~Torcini, Collective dynamics in sparse
  networks, Phys. Rev. Lett. 109 (2012) 138103.

\bibitem{monteforte2010dynamical}
M.~Monteforte, F.~Wolf, Dynamical entropy production in spiking neuron networks
  in the balanced state, Phys. Rev. Lett. 105~(26) (2010) 268104.

\bibitem{vanVreeswijk1996}
C.~van Vreeswijk, H.~Sompolinsky, Chaos in neuronal networks with balanced
  excitatory and inhibitory activity, Science 274~(5293) (1996) 1724--1726.

\bibitem{BenettinLyapunov1980}
G.~Benettin, L.~Galgani, A.~Giorgilli, J.-M. Strelcyn, Lyapunov characteristic
  exponents for smooth dynamical systems and for hamiltonian systems; a method
  for computing all of them. part 1: Theory, Meccanica 15~(1) (1980) 9--20.

\bibitem{LetzKantz2000}
T.~Letz, H.~Kantz, Characterization of sensitivity to finite perturbations,
  Phys. Rev. E 61 (2000) 2533--2538.

\bibitem{Cencini2010chaos}
M.~Cencini, F.~Cecconi, A.~Vulpiani, Chaos: From Simple Models to Complex
  Systems, Series on advances in statistical mechanics, World Scientific, 2010.

\bibitem{FSPolitiLE}
A.~Politi, Lyapunov exponent 8~(3) (2013) 2722.

\bibitem{Torcini1995FSLE}
A.~Torcini, P.~Grassberger, A.~Politi, Error propagation in extended chaotic
  systems, J. Phys. A-Math. Gen. 28~(16) (1995) 4533.

\bibitem{shibata1998}
T.~Shibata, K.~Kaneko, Collective chaos, Phys. Rev. Lett. 81 (1998) 4116--4119.

\bibitem{cencini1999}
M.~Cencini, M.~Falcioni, D.~Vergni, A.~Vulpiani, Macroscopic chaos in globally
  coupled maps, Physica D 130~(1) (1999) 58--72.

\bibitem{Olmi2010Chaos}
S.~Olmi, A.~Politi, A.~Torcini, {Collective chaos in pulse-coupled neural
  networks}, Europhys. Lett. (2010) 60007+.

\bibitem{CenciniTorcini}
M.~Cencini, A.~Torcini, Linear and nonlinear information flow in spatially
  extended systems, Phys. Rev. E 63 (2001) 056201.

\bibitem{vanVreeswijk1996Partial}
C.~van Vreeswijk, {Partial synchronization in populations of pulse-coupled
  oscillators}, Phys. Rev. E 54~(5) (1996) 5522--5537.

\bibitem{pikovsky1998dynamic}
A.~Pikovsky, A.~Politi, Dynamic localization of lyapunov vectors in spacetime
  chaos, Nonlinearity 11~(4) (1998) 1049.

\bibitem{politi2010lif}
A.~Politi, S.~Luccioli, Dynamics of networks of leaky integrate-and-fire
  neurons, in: Network Science: Complexity in Nature and Technology, Springer,
  London, 2010, p. 217.

\bibitem{Koch1998}
C.~Koch, Biophysics of Computation: Information Processing in Single Neurons
  (Computational Neuroscience), 1st Edition, Oxford University Press, 1998.

\bibitem{Sterratt2011Principles}
D.~Sterratt, B.~Graham, A.~Gillies, D.~Willshaw, {Principles of Computational
  Modelling in Neuroscience}, 1st Edition, Cambridge University Press, 2011.

\bibitem{mccormick1985comparative}
D.~A. McCormick, B.~W. Connors, J.~W. Lighthall, D.~A. Prince, Comparative
  electrophysiology of pyramidal and sparsely spiny stellate neurons of the
  neocortex, J. Neurophysiol. 54~(4) (1985) 782--806.

\bibitem{destexhe2001fluctuating}
A.~Destexhe, M.~Rudolph, J.-M. Fellous, T.~J. Sejnowski, Fluctuating synaptic
  conductances recreate in vivo-like activity in neocortical neurons,
  Neuroscience. 107~(1) (2001) 13--24.

\end{thebibliography}

\end{document}